\documentclass{article}

\usepackage[english]{babel}

\usepackage[letterpaper,top=1.5cm,bottom=2cm,left=1.5cm,right=1.5cm,marginparwidth=1.75cm]{geometry}

\usepackage{wrapfig}
\usepackage{amsmath}
\usepackage{graphicx}
\usepackage[colorinlistoftodos]{todonotes}
\usepackage[colorlinks=true, allcolors=blue]{hyperref}
\usepackage{caption}
\usepackage{subcaption}
\usepackage{sectsty}
\usepackage{float}
\usepackage{titling} 
\usepackage{blindtext}
\usepackage{units}
\usepackage{natbib}
\usepackage{xcolor}
\definecolor{darkgreen}{rgb}{0.0, 0.4, 0.0}
\usepackage{minted}
\usepackage{pdfpages}

\usepackage[table]{xcolor}   % for \rowcolors
\usepackage{booktabs}        % optional but nicer rules
\usepackage{multirow}
\usepackage{pdflscape}
\usepackage{longtable}
\usepackage{booktabs}

\usepackage{xcolor} % for text color

  {} % end: nothing special

\title{\textbf{State transitions in land–vegetation systems emerge at Paris Agreement warming levels in CMIP6}}
\author{
Pouriya Alinaghi$^{1,*}$,
Sybren Drijfhout$^{1,2,3}$,
Joran Angevaare$^{1}$,\\
Chris Jones$^{4}$,
Andy Wiltshire$^{4}$
\\[0.5em]
{\small $^{1}$Department of R\&D Weather and Climate Models, Royal Netherlands Meteorological Institute (KNMI), de Bilt, The Netherlands}\\
{\small $^{2}$Institute for Marine and Atmospheric Research, Utrecht University, Utrecht, The Netherlands
}\\
{\small $^{3}$Ocean and Earth Science, National Oceanography Centre Southampton, University of Southampton, Southampton, UK
}\\
{\small $^{4}$Met Office Hadley Centre, Exeter, UK}\\[0.5em]
{\small $^{*}$Corresponding author: Pouriya Alinaghi (\texttt{pouriya.alinaghi@knmi.nl})}
}

\begin{document}
\maketitle

% \tableofcontents

\noindent

%% 248 word now.
% \newpage
\begin{abstract}
Using an automatic detection workflow applied to the Coupled Model Intercomparison Project Phase 6 (CMIP6) ensemble under future emission scenarios, we identify 47 abrupt and more gradually developing state transitions in the land-vegetation component of the Earth system, classified into 9 categories. Over the Amazon, we find state transitions via vegetation dieback alongside greening cases; the contrast between them is traced primarily to differences in precipitation: models showing dieback experience either a larger absolute decline in precipitation, or one that translates more efficiently into soil moisture loss, particularly near the surface, while in greening models the CO$_2$ fertilization effect wins out where the soil moisture loss remains weaker. The precipitation decline in dieback-prone models appears driven through a weakening of moist convection. Across these Amazon cases, models with dynamic vegetation undergo dieback, whereas greening is confined to models with prescribed vegetation distributions. African cases include greening over eastern-central Africa and the Congo basin, and an abrupt soil-moisture drying also over the Congo. At high latitudes, boreal forest expands, while permafrost thaws once the regional above-zero temperatures persist for more than half the year. Additional categories cover transitions to a reduced snow-cover state over northeastern North America, increased vegetation biomass near the Tibetan Plateau and southeastern Asia, and increased leaf-area index over the northeast Northern America. Of particular concern, a global warming of 2$^\circ$C or below, within reach of the Paris Agreement targets, is already enough to trigger the onset of the majority of the identified categories in CMIP6.
\end{abstract}

\section{Introduction}

Land–vegetation components play a central role in the Earth's carbon cycle and climate system. The Amazon rainforest — the world's largest tropical forest — stores approximately 150–200\, gigatonnes of carbon (GtC) in its biomass \citep[e.g.,][]{gatti2021amazonia,nobre2021amazon,mckay2022tipping}, while permafrost soils across the high latitudes of the Northern Hemisphere lock away an estimated 1035\,GtC in frozen ground \citep[e.g.,][]{schuur2015climate, mckay2022tipping}. Both systems are under growing pressure: deforestation directly degrades the Amazon's carbon sink capacity, while anthropogenically driven global warming threatens the structural integrity of both the forest and permafrost. Under continued warming, these systems risk crossing thresholds beyond which they transition from carbon sinks to carbon sources, releasing stored carbon back into the atmosphere, accelerating the rise in atmospheric CO$_2$ concentrations, and thereby further amplifying global warming \cite[e.g.,][]{cox2000carbon,turetsky2020perma} with potentially severe consequences for the global climate \citep[e.g.,][]{brovkin2021past, mckay2022tipping}.

%%% transition to tipping element paragraph
Such systems are known as \textit{tipping elements} of the climate system \citep{lenton2008tipping}. A tipping element is a component of the Earth system that, once a critical forcing threshold is crossed, undergoes a transition driven predominantly by internal self-reinforcing feedbacks rather than by the external forcing itself. These feedbacks propel the system toward a qualitatively different state at a pace that far exceeds the rate of the forcing — changes that may be abrupt \cite[i.e. a few decades or less according to AR6, IPCC, Chapter 4,][]{lee2021ar6ts} — and critically, the transition is not reversed when the forcing is reduced, implying irreversibility, at least on human timescales \citep[e.g.,][]{scheffer2001catastrophic, hirota2011amazon, lenton2008tipping}.

%%% also on tipping
These characteristics have raised growing concern among scientists and policymakers alike. Beyond successive IPCC reports \citep{ipcc2021ar6wg1}, this concern has given rise to dedicated Global Tipping Points Reports \citep{lenton2023tipping, lenton2025tipping} and a body of expert-elicitation studies estimating the warming levels at which specific systems may undergo drastic change \citep{mckay2022tipping, loriani2025tipping}. It should be noted, however, that the tipping point framing \citep{lenton2008tipping} has itself faced criticism: some argue that casting dangerous climate change in terms of discrete thresholds may promote passivity in mitigation and adaptation, given that impacts are already severe and accumulating continuously across all warming levels \citep{kopp2025tipping}. Toward a more systematic and objective characterization of these transitions, \cite{drijfhout2015abrupt} produced a catalogue of abrupt shifts across the Coupled Model Intercomparison Project Phase 5 \citep[CMIP5;][]{taylor2012cmip5} ensemble through manual inspection of numerous Earth system variables. Building on this, \cite{bathiany2020tipping} automated shift detection in CMIP5 by applying the Canny edge-detection algorithm \citep{canny1986edge}, an image-processing technique that identifies sharp gradients in time and space, and \cite{terpstra2025tipping} subsequently extended this approach to the idealized 1\%\,yr$^{-1}$ CO$_2$ increase (1pctCO2) scenario of CMIP6 \citep{eyring2016cmip6}. Focused specifically on Amazon vegetation under the same idealized scenario, \cite{parry2025amazon} found that the area affected by localized dieback in northern South America increased by 2--12\% per degree of warming beyond 1.5$^\circ$C.

%%% gap and this study
Together, these studies have substantially advanced the detection of abrupt nonlinear shifts in Earth system models, yet two important gaps remain. First, all of the above focus exclusively on abrupt transitions, leaving more gradually developing but potentially consequential state transitions in land--vegetation components largely uncharacterized. Second, the 1pctCO2 scenario — while useful for process understanding — does not capture the socio-economic diversity and forcing trajectories of more realistic emission pathways. The CMIP6 ensemble addresses this through its Shared Socioeconomic Pathway (SSP) scenarios \citep{scenariomip2016cmip6, eyring2016cmip6}, which span a wide range of CO$_2$ forcing magnitudes, rates, and associated warming levels. Here, we fill both gaps simultaneously by applying the workflow of \cite{angevaare2025catalogue} — previously developed and validated for physical ocean and sea-ice components \citep{angevaare2025catalogue} — to the land--vegetation components of CMIP6 under the SSP1-2.6, SSP2-4.5, SSP3-7.0, and SSP5-8.5 scenarios \citep{scenariomip2016cmip6}. This approach detects both abrupt and more gradually developing state transitions — collectively termed Strong Nonlinear Surprises~(SNSs) — in a largely objective manner, without relying on expert judgment to preselect regions or variables of interest, representing a methodological advance over studies such as \cite{drijfhout2015abrupt}, \cite{mckay2022tipping}, and \cite{loriani2025tipping}. This study makes three contributions. First, we catalogue SNSs in land--vegetation components across a policy-relevant range of forcing pathways, allowing us to assess how the timing, magnitude, and frequency of nonlinear transitions depend on the level of future warming. Second, for each identified SNS, we qualitatively investigate and propose the underlying physical and ecological mechanisms, moving beyond detection toward process understanding and providing a basis for targeted future studies. Third, by anchoring SNSs to global warming levels, we provide multi-model estimates of the warming thresholds at which land--vegetation systems may undergo abrupt or more gradually developing state shifts, complementing existing assessments based on expert judgments or the idealized 1pctCO2 scenario \citep[e.g.,][]{mckay2022tipping, loriani2025tipping, terpstra2025tipping}.

%%% paper structure
The remainder of this paper is structured as follows. Section~\ref{sec: data and methods} describes the CMIP6 variables and models used, explains how the \cite{angevaare2025catalogue} workflow is adapted for land–vegetation components, and details how the identified SNSs are grouped, presented, and anchored to global warming level distributions. In Section~\ref{sec: results}, we present 47 SNS cases organized into nine categories spanning the Amazon rainforest, central Africa, the boreal zone, Arctic permafrost, northern hemisphere snow cover, the Tibetan Plateau and East Asia, and eastern North America, discuss the physical and ecological mechanisms driving each transition, and identify the global warming levels at which SNSs emerge and levels at which undergo their steepest rate of change. Section~\ref{sec:discussion} summerizes the results and discusses methodological limitations and the representation of land–vegetation dynamics in CMIP6, before closing with implications for the terrestrial carbon cycle, ecosystem services, and the populations whose livelihoods and food security depend on these regions.

% \newpage
\section{Data \& Methodology}
\label{sec: data and methods}
\subsection{Detection of strong nonlinear surprises in land-based components of climate model simulations}
We apply the automated Strong Nonlinear Surprise (SNS) detection workflow of \cite{angevaare2025catalogue} to land-based components of the climate system. Specifically, the workflow is applied to continuous simulations formed by concatenating the historical experiment with each shared socio-economic pathway scenario (SSP1-1.9, SSP1-2.6, SSP2-4.5, SSP3-7.0, and SSP5-8.5), across all available ensemble members from 54 climate models participating in CMIP6. The analysis is performed on two-dimensional, re-gridded, yearly averaged fields of near-surface air temperature (tas), sea-level pressure (psl), soil moisture (mrso), precipitation (pr), evapotranspiration (evspsbl), total cloud fraction (clt), snow-area fraction (snc), leaf-area index (lai), tree-area fraction (treeFrac), grass-area fraction (grassFrac), and bare-soil area fraction (baresoilFrac). Owing to differences in the variables and ensemble members each model provided through the Earth System Grid Federation (ESGF) nodes at the time of this study, per-model coverage is uneven; summed over all models, variables, and scenarios, this yields approximately 11,500 analyzed two-dimensional fields ($\sim$1{,}800 distinct model--scenario--member simulations), with CanESM5, EC-Earth3, MIROC6, and MPI-ESM1-2-LR contributing the most to the total data with their large number of ensemble members (see Table~\ref{tab: all available data}).

The SNS identification proceeds in three successive steps, briefly summarized here. In the first step, the algorithm isolates candidate regions, i.e., spatially coherent areas with a high likelihood of exhibiting an SNS. We follow the same method as \cite{angevaare2025catalogue}, which yields regions of at least 10$^6$~km$^2$. This region-isolation step is intentionally permissive: it is designed to reduce computational complexity and therefore favors the inclusion of potentially relevant regions over the exclusion of true SNS candidates.

In the second step, the spatially averaged time series of each candidate region is tested against a set of formal criteria that distinguish two classes of SNS: abrupt changes and state transitions. Abrupt changes are rapid shifts that exceed a prescribed magnitude over decadal timescales, for example a variable that jumps to a distinctly different level within one or two decades and remains there. State transitions, by contrast, are sustained departures from the prior state that are too large to be explained by external forcing alone and thus point to strong internal feedbacks; they need not be rapid, but they leave the system in a qualitatively different regime. When a region satisfies more than one criterion, the corresponding regions are merged, provided the merged region still meets the criteria, so that SNS are not smoothed out through aggregation.

Of the six criteria defined by \cite{angevaare2025catalogue}, three apply to the land and vegetation variables considered here: two identify abrupt changes (their criteria i and ii) and one identifies state transitions (their criterion vi). Criterion (i) flags an abrupt change when the largest decadal jump in the smoothed time series is large relative to internal (piControl) variability and the time series is clearly bimodal, i.e., the system occupies two distinct states before and after the shift. Criterion (ii) captures the same kind of abrupt change when it occurs near the end of the simulation, where the post-transition state is too short to establish bimodality; it therefore applies the same magnitude thresholds but relaxes the bimodality requirement. Criterion (vi) targets the more gradually developing state transitions: it requires that the overall start-to-end change be large both relative to internal decadal variability and relative to the forced (zonal-mean) change, so that the signal reflects an internally driven shift rather than the externally forced trend. Their remaining criteria (iii)--(v) are specific to ocean mixed-layer depth, ocean meridional circulation, and sea ice and are therefore not applicable here. We refer the reader to the method paper \cite[see their appendix, ][]{angevaare2025catalogue} for the exact threshold values and the full algorithm.

In the last step, individual SNS time series are first combined into cases, where a case groups all SNS time series originating from the same model and occurring over spatially overlapping regions, irrespective of the scenario, ensemble member, or variable through which they are detected. Cases are subsequently grouped into categories according to their geographic location and (possibly) the shared physical mechanism underlying the transition. For each individual SNS time series, the global warming levels at which it emerges as well as the levels at which it undergoes its most rapid transition are estimated using the flexible window approach detailed in Section~\ref{sec:methods_warming}, following \citet{angevaare2025catalogue}. These per-time-series distributions are then aggregated progressively from the case level to the category level, as described in Section~\ref{sec:methods_warming}.

This study is concerned with strong nonlinear \textit{surprises}. Land-use forcing in the SSP scenarios, by construction, is a prescribed rather than emergent change, and therefore does not constitute a surprise in this sense. The SNS detection algorithm of \citet{angevaare2025catalogue}, however, identifies transitions based on their variations in space and time without taking into account their physical origin, and consequently also flagged many cases driven purely by land-use forcing — for instance, the conversion of forest to cropland or pasture. We manually inspected these cases and removed those attributable solely to land-use forcing from the analysis. If a case is inspected to be a mixture of both land-use forcing and natural response of the system, we present the case here and gauge the scenario simulation against the quadrupling CO$_2$ (4xCO2) scenario to disentangle the natural response of the system from the land-use forcing.

\subsection{Global warming level calculations for when an SNS emerges and reaches it maximum change}
\label{sec:methods_warming}

To characterize the probability distribution function (PDF) of the warming level at which each SNS undergoes its most rapid transition, we adopt the flexible time window methodology of \citet{angevaare2025catalogue}. We additionally determine the warming level at which each transition \textit{begins}, developed here. Both metrics are obtained through the same three steps.

\textit{Step 1: Individual time series PDF.} For a single SNS time series (corresponding to one model, one scenario, one ensemble member, and one variable; e.g., the gradual decline in leaf-area index in IPSL-CM6A-LR under SSP5-8.5, member r1i1p1f1), we identify the year at which the change over a sliding window of $w$ years in the $w$-year running mean is maximal, and assign equal probability to all years within that window. This is repeated for $w$ ranging from 20 to 50 years, producing a distribution of years over which the steepest change is most likely to occur. Since each year has a corresponding global mean surface temperature anomaly relative to the 1850--1880 baseline, this year-probability distribution is directly translated into a probability distribution over warming levels $\Delta T$ \cite[see their Fig.~A1,][]{angevaare2025catalogue}. The result is a PDF, $p_{m,s,e,v}(\Delta T)$, describing the probability that the steepest change occurs at warming level $\Delta T$ for model $m$, scenario $s$, ensemble member $e$, and variable $v$.

Since the rate of change necessarily rises from near zero to its maximum, the onset of a transition occurs, by construction, at an earlier year and most likely at a lower warming level than the steepest-change level defined above. It is obtained within this same first step; only the year-identification changes, while Steps~2 and~3 below are applied unchanged, with the onset-based PDF derived below substituted for $p_{m,s,e,v}(\Delta T)$ throughout. For a single SNS time series, smoothed with a $w$-year running mean, we fit two candidate piecewise-linear models: one describing a single transition between two linear regimes,
\begin{equation}
    y(t) =
    \begin{cases}
        a_1\,t + b_1, & t \le t_1\\
        a_2\,t + b_2, & t > t_1,
    \end{cases}
\end{equation}
and one additionally allowing for a return to a stable regime once the transition is complete,
\begin{equation}
    y(t) =
    \begin{cases}
        a_1\,t + b_1, & t \le t_1\\
        a_2\,t + b_2, & t_1 < t \le t_2\\
        a_3\,t + b_3, & t > t_2,
    \end{cases}
\end{equation}
with continuity enforced at the breakpoint $t_1$ and, where present, the second breakpoint $t_2$. Here $t_1$ and $t_2$ are not selected from a predefined set of candidate years, but are themselves free parameters of the fit: for each form, all parameters ($t_1,a_1,b_1,a_2,b_2$), or their five-parameter analogue including ($t_2,a_3,b_3$) are estimated jointly by nonlinear least-squares, minimizing the total squared difference between the piecewise function and the smoothed series. The fitted $t_1$, being continuous-valued, is rounded to the nearest calendar year in the series to obtain the onset year used below.

Because the three-segment form has more free parameters, it can always match the data at least as well as the two-segment form, even when the third regime reflects noise rather than a genuine second change in behaviour; fit quality alone therefore cannot decide between them. We instead compare the two forms using the Akaike Information Criterion, AIC \citep{akaike1974new}, which rewards a lower residual sum of squares (RSS) but explicitly penalizes each additional free parameter,
\begin{equation}
    \mathrm{AIC} = n\ln(2\pi) + n + n \ln\!\left(\frac{\mathrm{RSS}}{n}\right) + 2k,
\end{equation}
where $n$ is the number of time steps and $k$ the number of free parameters (four or six, for the two- and three-segment forms respectively). The three-segment form is retained only when it reduces the AIC by more than 2 relative to the two-segment form -- a conventional threshold for judging the improvement large enough to justify the extra complexity -- and the two-segment form is used otherwise. A fit is accepted only if every breakpoint lies strictly within the time series and the transition slope differs clearly from its neighboring regime(s) ($|a_2-a_1|>2|a_1|$, and $|a_2-a_3|>2|a_3|$ where a third regime is present); this discards cases where the series is effectively linear over the fitted window, while correctly treating a reversal in the sign of the slope -- not only an increase in its magnitude -- as clear evidence of a transition. In practice, this criterion discards none of the SNS time series in our catalogue, since every case classified as an SNS must, by definition, already depart from linear behaviour.
The onset year is identified as $t_1$, the point at which the series departs from its background trend into the transition. As for the steepest-change metric above, this is repeated for $w$ ranging from 20 to 50 years, with equal probability assigned to all years within the $w$-year window centered on $t_1$, and converted to a warming-level PDF, $p^{\mathrm{onset}}_{m,s,e,v}(\Delta T)$, via the same year-to-temperature mapping used above. Two representative examples of the fitting procedure, one requiring only two regimes and one requiring three, are given in the supplement (Fig.~\ref{fig: example fits}).

\textit{Step 2: Case-level PDF.} A single SNS case (e.g., the Amazon dieback in IPSL-CM6A-LR) corresponds to one model-region (e.g., IPSL-CM6A-LR, Amazon) but may be detected across multiple scenarios, ensemble members, or variables (e.g., could be detected by both tree fraction and leaf-area index and in multiple members or scenarios). The case-level PDF is obtained by averaging the PDFs of individual SNSs equally over all contributing combinations:
\begin{equation}
    p_{\mathrm{case}}(\Delta T) =
    \frac{1}{N_s N_e N_v} \sum_{s} \sum_{e} \sum_{v} p_{m,s,e,v}(\Delta T),
\end{equation}
where $N_s$, $N_e$, and $N_v$ denote the number of contributing scenarios, ensemble members, and variables, respectively. This averaging combines all detections of a case into a single case-level PDF, so that a case detected in more scenarios, ensemble members, or variables is not over-represented in the category-level step that follows.

\textit{Step 3: Category-level PDF.} A category groups multiple cases that share the same geographic location and possibly the same underlying physical mechanism, each from a different model. The category-level distribution is obtained by averaging the case-level PDFs over all $N_{\mathrm{cases}}$ contributing cases:
\begin{equation}
    p_{\mathrm{cat}}(\Delta T) = \frac{1}{N_{\mathrm{cases}}}
    \sum_{c=1}^{N_{\mathrm{cases}}} p_{\mathrm{case},c}(\Delta T),
\end{equation}
so that each case -- and hence each model -- contributes equally to the category-level distribution. The resulting category-level distributions are shown in Section~\ref{sec:warming_levels}. Full details of the method and its implementation are given in the method paper \cite[see their appendix,][]{angevaare2025catalogue}.

\subsection{Moisture budget analysis for atmosphere and land}

Many of the SNS elements in this study occur over rainforest regions, including the Amazon and Congo basins. These systems are strongly coupled to the hydrological cycle across both land and atmosphere. To characterize this coupling and interpret the behaviour of each model and case, we perform a mass-weighted vertically integrated moisture budget analysis. For the atmosphere, the moisture budget is given by
\begin{align}
    \partial_t I_a = \partial_t \langle q_t \rangle = \text{Evapotranspiration} - \text{Precipitation} - \underbrace{\langle \nabla \cdot (\mathbf{u} q_t) \rangle}_{\text{Horizontal moisture flux divergence}},
\end{align}
where $\langle \cdot \rangle$ denotes mass-weighted vertical integration over the full atmospheric column, $I_a$ is the mass-weighted vertically integrated atmospheric moisture content, $q_t$ is the total specific humidity, and $\mathbf{u} = (u_x, u_y)$ represents the horizontal wind vector in the east--west and north--south directions. Similarly, the moisture budget for land is written as
\begin{align}
    \partial_t I_s = \partial_t \langle q_s \rangle = \text{Precipitation} - \text{Evapotranspiration} - \text{Runoff},
\end{align}
where $I_s$ is the vertically integrated soil water content and $q_s$ denotes soil moisture.

In the atmosphere, evapotranspiration and precipitation act as a source and sink of moisture, respectively, while horizontal moisture flux divergence represents either a source (convergence) or a sink (divergence), depending on the large-scale circulation. Over land, precipitation and evapotranspiration act as source and sink, respectively, while runoff represents lateral and gravitational drainage of water from the soil column.

For all models analyzed in this study, these budget terms are evaluated using monthly mean variables. Precipitation (\text{pr}), evapotranspiration (\text{evspsbl}), and total runoff (\text{mrro}) are available for most models. However, the mass-weighted vertically integrated moisture transport terms $u_x q_t$ (intuaw) and $u_y q_t$ (intvaw) are only available for a subset of models, limiting the completeness of the atmospheric moisture budget analysis. Furthermore, we do not attempt to close the moisture budgets, as offline computation of the tendencies due to spatial divergence of transport terms from monthly mean output introduces substantial numerical errors \cite[e.g., see][]{seager2013diagnostic}, which accumulate over the multi-century simulations considered here. Nevertheless, for each case we analyze the total soil moisture $I_s$ (or for some specific cases also $I_a$), which provides a clear indication of how the individual moisture budget terms influence the overall moisture state of the soil (and the atmosphere).

% \newpage
\section{Results}
\label{sec: results}
Applying the workflow results in the identification of 47 SNS cases classified into 9 categories (Table~\ref{tab: all SNS cases}). The first two categories concern the Amazon rainforest: two cases project a state transition of the Amazon basin via forest dieback (category A$-$, Section~\ref{sec: grad amazon dieback}), one case projects an abrupt Amazon dieback (category a, Section~\ref{sec: abrupt amazon dieback}), and three cases project Amazon greening (category A$+$, Section~\ref{sec: amazon greening}). The 3rd and 4th categories cover central Africa, with five cases projecting greening (category B, Section~\ref{sec: africa greening - 1}) and one case projecting an abrupt decline in soil moisture over the Congo basin (category b, Section~\ref{sec: africa drying}). The 5th category captures boreal forest expansion across six cases (category C, Section~\ref{sec: boreal forest}). The 6th category encompasses permafrost collapse (categories D and d, Section~\ref{sec: permafrost}), with 12 state transitions and 5 abrupt cases respectively. The 7th category comprises two cases of transitions of the northeast North America to a low snow-cover state (category E, Section~\ref{sec: snow decline}). The 8th category covers greening across multiple regions of the Tibetan Plateau and East Asia in seven cases (category F, Section~\ref{sec: asia greening}). Finally, the 9th category consists of three cases projecting greening over eastern North America (category G, Section~\ref{sec: usa greening}). For all the aforementioned categories, the global warming levels at which their corresponding SNSs experience their onset and steepest rate of change are presented in Section~\ref{sec:warming_levels}; it shows that the land-vegetation systems feature SNSs that already begin to unfold at warming levels within the range of Paris agreement target.

The remainder of this section examines each category in turn. For each category, we present SNS cases through their time series and spatial extent, with at least one example shown per case. Where multiple ensemble members produce an SNS within the same case, we show the member with the largest affected area. We then briefly characterize the state of the relevant land-surface and atmospheric variables --- including leaf-area index, soil moisture, tree cover fraction, snow-cover area, precipitation, and evapotranspiration --- to provide context for the hydrological and climatic conditions under which each SNS emerges. Each category closes with a qualitative discussion of the physical mechanisms responsible for the transition, supported by the accompanying figures.

%%% overview of all cases
\begin{table}[t!]
\centering
\tiny
\begin{tabular}{rllllll}
\toprule
\textbf{Case} & \textbf{Category} &         \textbf{Model} &                            \textbf{SSP Scenario} & \textbf{Criterion} &                        \textbf{Variable} &            \textbf{Region} \\
\midrule
    1 &       A$+$ &   CESM2-WACCM &                 370, 585 &   vi &                         lai &   N.South-America \\
    2 &       A$+$ &         CESM2 &                 370, 585 &   vi &                         lai &   N.South-America \\
    3 &       A$+$ &    NorESM2-LM &                 245, 585 &   vi &                         lai &   N.South-America \\
    \midrule
    4 &       A$-$ &  IPSL-CM6A-LR &                         585 &   vi &                         lai &   N.South-America \\
    5 &       A$-$ &   UKESM1-0-LL &                         585 &   vi & baresoilFrac, lai, treeFrac &   N.South-America \\
    % \midrule
    6 &        a &     GFDL-ESM4 &                         370 &    i &                    treeFrac &   N.South-America \\
    \midrule
    7 &        B &     CanESM5-1 &                         585 &   vi &                         lai &    Central-Africa \\
    8 &        B & CanESM5-CanOE &                 370, 585 &   vi &                         lai &    Central-Africa \\
    9 &        B &       CanESM5 &         245, 370, 585 &   vi &                         lai &    Central-Africa \\
   10 &        B &    NorESM2-LM &                         245 &   vi &                         lai &    Central-Africa \\
   11 &        B &   UKESM1-0-LL &                         585 &   vi &                baresoilFrac &  N.Eastern-Africa \\
   \midrule
   12 &        b &    MRI-ESM2-0 &                         585 &    i &                        mrso &    Central-Africa \\
   \midrule
   13 &        C &     CanESM5-1 &                         585 &   vi &                         lai &         W.Siberia \\
   14 &        C & CanESM5-CanOE &         245, 370, 585 &   vi &                         lai &         W.Siberia \\
   15 &        C &       CanESM5 &         245, 370, 585 &   vi &                         lai &         W.Siberia \\
   16 &        C &       CanESM5 &                         585 &   vi &                         lai & N.W.North-America \\
   17 &        C &   UKESM1-0-LL &         126, 370, 585 &   vi &      baresoilFrac, treeFrac &    Russian-Arctic \\
   18 &        C &   UKESM1-0-LL &                         585 &   vi &                    treeFrac & N.W.North-America \\
   \midrule
   19 &        D & CNRM-CM6-1-HR & 126, 245, 370, 585 &   vi &                        mrso & N.E.North-America \\
   20 &        D & CNRM-CM6-1-HR &                         585 &   vi &                        mrso &  Russian-Far-East \\
   21 &        D &    CNRM-CM6-1 &         245, 370, 585 &   vi &                        mrso & N.E.North-America \\
   22 &        D &    CNRM-CM6-1 &                         585 &   vi &                        mrso & N.W.North-America \\
   23 &        D &    CNRM-CM6-1 &                         585 &   vi &                        mrso &    Russian-Arctic \\
   24 &        D &   CNRM-ESM2-1 &                 370, 585 &   vi &                        mrso &         E.Siberia \\
   25 &        D &   CNRM-ESM2-1 &                         585 &   vi &                        mrso & N.W.North-America \\
   26 &        D &    NorESM2-LM &                         585 &   vi &                        mrso & N.E.North-America \\
   27 &        D &    NorESM2-MM &                         585 &   vi &                        mrso & N.W.North-America \\
   28 &        D &    NorESM2-MM &                         585 &   vi &                        mrso &    Russian-Arctic \\
   29 &        D &    NorESM2-MM &                         585 &   vi &                        mrso &    Russian-Arctic \\
   30 &        D &       TaiESM1 &                 370, 585 &   vi &                        mrso &    Russian-Arctic \\
   % \midrule
   31 &        d &    ACCESS-CM2 &                         585 &    i &                        mrso & N.E.North-America \\
   32 &        d & ACCESS-ESM1-5 &                         585 &    i &                        mrso & N.E.North-America \\
   33 &        d &      E3SM-1-1 &                         585 &    i &                        mrso & N.E.North-America \\
   34 &        d &   UKESM1-0-LL &                 370, 585 &    i &                        mrso &    Russian-Arctic \\
   35 &        d &   UKESM1-0-LL &                 370, 585 &    i &                        mrso & N.W.North-America \\
   \midrule
   36 &        E &   GISS-E2-1-G &                         585 &   vi &                         snc & N.E.North-America \\
   37 &        E &    NorESM2-MM &                         585 &   vi &                         snc & N.E.North-America \\
   \midrule
   38 &        F &   CESM2-WACCM &                         585 &   vi &                         lai &   Tibetan-Plateau \\
   39 &        F &         CESM2 &                         585 &   vi &                         lai &   Tibetan-Plateau \\
   40 &        F &     CanESM5-1 &                         585 &   vi &                         lai &            E.Asia \\
   41 &        F & CanESM5-CanOE &                 370, 585 &   vi &                         lai &            E.Asia \\
   42 &        F &       CanESM5 &         245, 370, 585 &   vi &                         lai &            E.Asia \\
   43 &        F &    NorESM2-LM &                         585 &   vi &                         lai &            E.Asia \\
   44 &        F &   UKESM1-0-LL &                         585 &   vi &                baresoilFrac &         W.Siberia \\
   \midrule
   45 &        G &     CanESM5-1 &                         585 &   vi &                         lai &   E.North-America \\
   46 &        G & CanESM5-CanOE &                         585 &   vi &                         lai &   E.North-America \\
   47 &        G &       CanESM5 &                         585 &   vi &                         lai &   E.North-America \\
\bottomrule
\end{tabular}
\caption{List of all 47 SNS cases identified in this study. Uppercase and lowercase category letters denote state transitions and abrupt changes, respectively; A$+$ and A$-$ distinguish Amazon greening and dieback. Criterion abbreviations follow \citet{angevaare2025catalogue}: (i) abrupt change; (vi) more  gradually developing state transition. Variable abbreviations: lai (leaf-area index), baresoilFrac (bare soil fraction), treeFrac (tree cover fraction), mrso (total soil moisture content), snc (snow cover fraction). Table~\ref{tab: fraction of members giving SNS} reports the fraction of members qualified as SNS, per case, scenario and variable.}
\label{tab: all SNS cases}
\end{table}

\subsection{Amazon Basin}

\subsubsection{(A$-$) State transition of the Amazon basin into a low-vegetation state via forest dieback}
\label{sec: grad amazon dieback}

\begin{figure}[t!]
    \centering
    \includegraphics[width=0.7\linewidth]{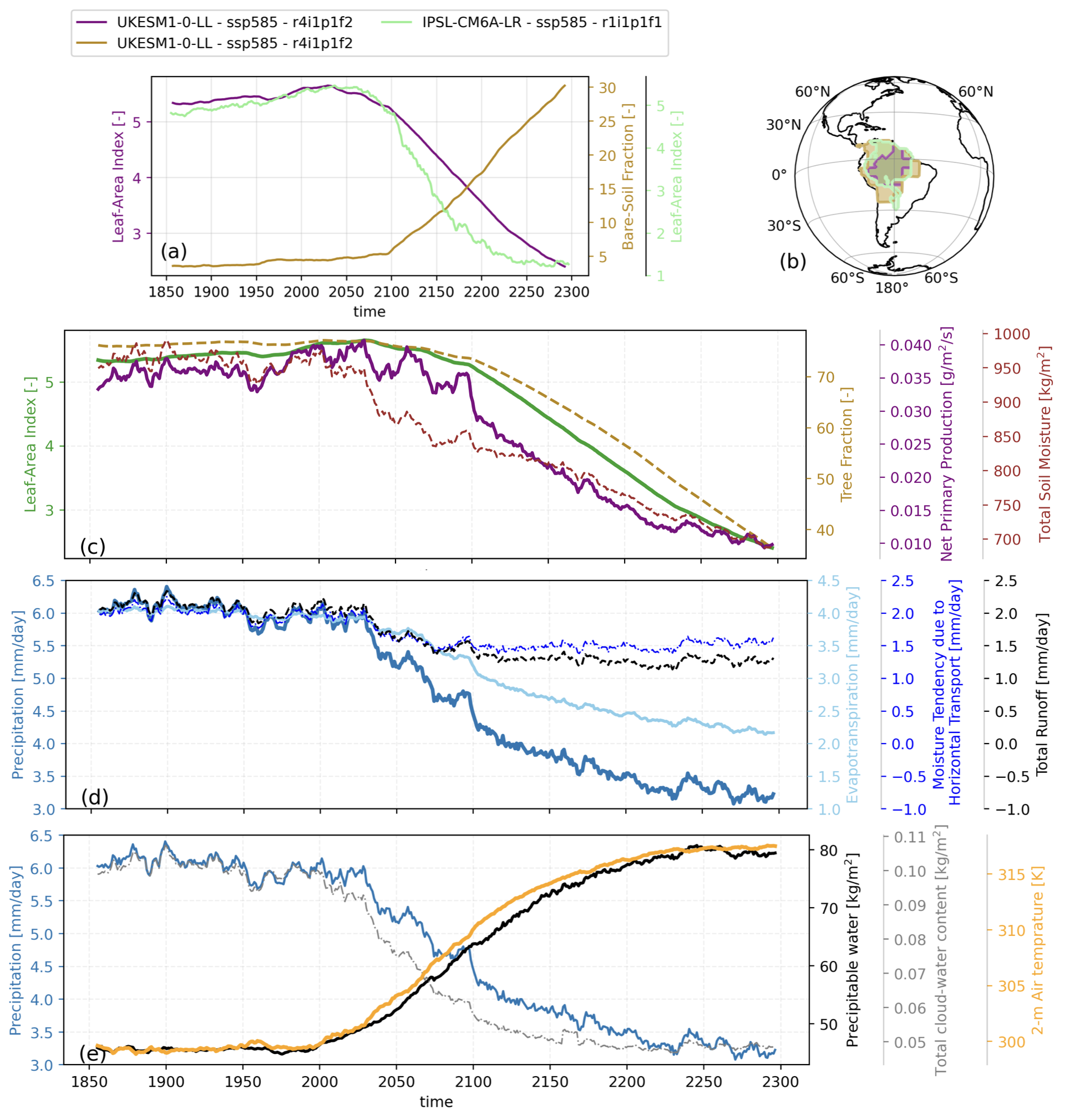}
    \caption{(a) Temporal evolution (10-year running mean) of leaf-area index and bare-soil fraction; corresponding spatial extent is shown in (b). (c–e) Time series of land and atmospheric variables, respectively, averaged over the region associated with the state transition (decline) in leaf-area index in the UKESM1-0-LL model (purple line in a).}
    \label{fig: Amazon - decline - 1}
\end{figure}

Two models, including one dynamic-vegetation model, UKESM1-0-LL (SSP5-8.5) and one non-dynamic vegetation model, IPSL-CM6A-LR (SSP5-8.5), project a transition of the Amazon basin into a low leaf-area index state (see Fig. \ref{fig: Amazon - decline - 1}a,b). In addition, the workflow identifies an increase in bare-soil fraction in the UKESM1-0-LL model as an SNS. For all the models, the changes in vegetation types such as pasture, crop and shrub are negligible, indicating that these cases are not due to the land-use change (not shown). As an example, we analyze the UKESM1-0-LL model, in which the reduction in leaf-area index is consistent with decreases in tree fraction and net primary production (Fig. \ref{fig: Amazon - decline - 1}c). Notably, the decline in net primary production begins around 2100, approximately 70 years after the initial reduction in soil moisture around 2030, suggesting that CO$_2$ fertilization under SSP5-8.5 temporarily compensates for the effect of soil moisture decline on vegetation productivity.

The soil moisture reduction is closely synchronized with a decline in precipitation around 2030, which the atmospheric moisture budget attributes to a concurrent reduction in horizontal moisture transport, suggesting an externally forced change in large-scale atmospheric circulation (Fig. \ref{fig: Amazon - decline - 1}d). At the land surface, the $\sim$0.8 mm/day precipitation reduction around 2030 is partly offset by a $\sim$0.4 mm/day decline in total runoff, but evapotranspiration does not adjust proportionally in this early phase likely due to sustained CO$_2$ fertilization, so the remaining imbalance depletes soil moisture. After 2050, as soil moisture crosses the vegetation stress threshold, net primary production and tree fraction decline, reducing evapotranspiration and further suppressing precipitation through moisture recycling — a feedback consistent with the strong evapotranspiration-precipitation coupling in this region and model, where evapotranspiration accounts for $\sim$65\% of precipitation in the historical period. After 2100, total runoff and horizontal moisture transport remain stable while precipitation, evapotranspiration and soil moisture decline together, confirming that the local vegetation-hydrology feedback is the dominant mechanism driving sustained forest dieback \citep[e.g.,][]{good2011forestextent}. 

Note that after around 2030, the total precipitation decline exceeds the sum of decline in evapotranspiration and transport-driven moisture tendency, implying a residual moistening of the atmospheric column. Indeed, the total precipitable water (water vapor content) over the region increases (Fig. \ref{fig: Amazon - decline - 1}e). However, this increased water-vapor content does not translate into increased cloud condensate (Fig. \ref{fig: Amazon - decline - 1}e). The strong regional warming, $\sim$18 K over the dieback region, substantially increases the atmosphere's water-holding capacity via Clausius-Clapeyron scaling, meaning a larger fraction of atmospheric moisture remains as vapor rather than condensing into cloud water (Fig. \ref{fig: Amazon - decline - 1}e). Notably, cloud-water content begins declining before the acceleration in precipitation loss (Fig. \ref{fig: Amazon - decline - 1}e). Collectively, these results indicate that precipitation efficiency — the fraction of atmospheric moisture that is converted into rainfall — is declining over the Amazon, suggesting a weakening of convective activity in this model, which could also be driven by the forest-dieback-induced decline in surface fluxes such as evapotranspiration. Accordingly, following the framework of \citet{held2006precip}, in which precipitation is controlled (i) thermodynamically, by moisture availability, and (ii) dynamically, via convective mass flux, our results suggest that the thermodynamically driven increase in precipitation over this region must be outweighed by a weakening of the dynamical component, presumably induced by reduced convective mass flux, resulting in a net precipitation decline (Fig. \ref{fig: Amazon - decline - 1}e; \citep{chadwick2013precip}). As convection is heavily parameterized in coarse-resolution Earth system models, the physical plausibility of such a decline in convective activity and precipitation efficiency remains an important open question that warrants dedicated process-level investigations.

The IPSL-CM6A-LR model exhibits a similar chain of processes to that in the UKESM1-0-LL model (Fig. \ref{fig: Amazon - decline - ipsl}): reduced moisture transport into the region appears to initialize a decline in precipitation (Fig. \ref{fig: Amazon - decline - ipsl}a,b), which in turn reduces soil moisture and evapotranspiration (Fig. \ref{fig: Amazon - decline - ipsl}a,b), thereby reinforcing the reduction in precipitation. Similarly, precipitation efficiency declines as total precipitable water over the region increases (Fig. \ref{fig: Amazon - decline - ipsl}c). Cloud-water content initially decreases in phase with the precipitation decline; however, after total precipitable water reaches approximately 75 kg/m$^2$ around 2100, cloud-water content begins recovering and stabilizes around 2200, coinciding with the leveling off of precipitable water (Fig. \ref{fig: Amazon - decline - ipsl}c).

\subsubsection{(a) Abrupt decline in tree fraction over Amazon basin}
\label{sec: abrupt amazon dieback}

\begin{figure}[t!]
    \centering
    \includegraphics[width=0.7\linewidth]{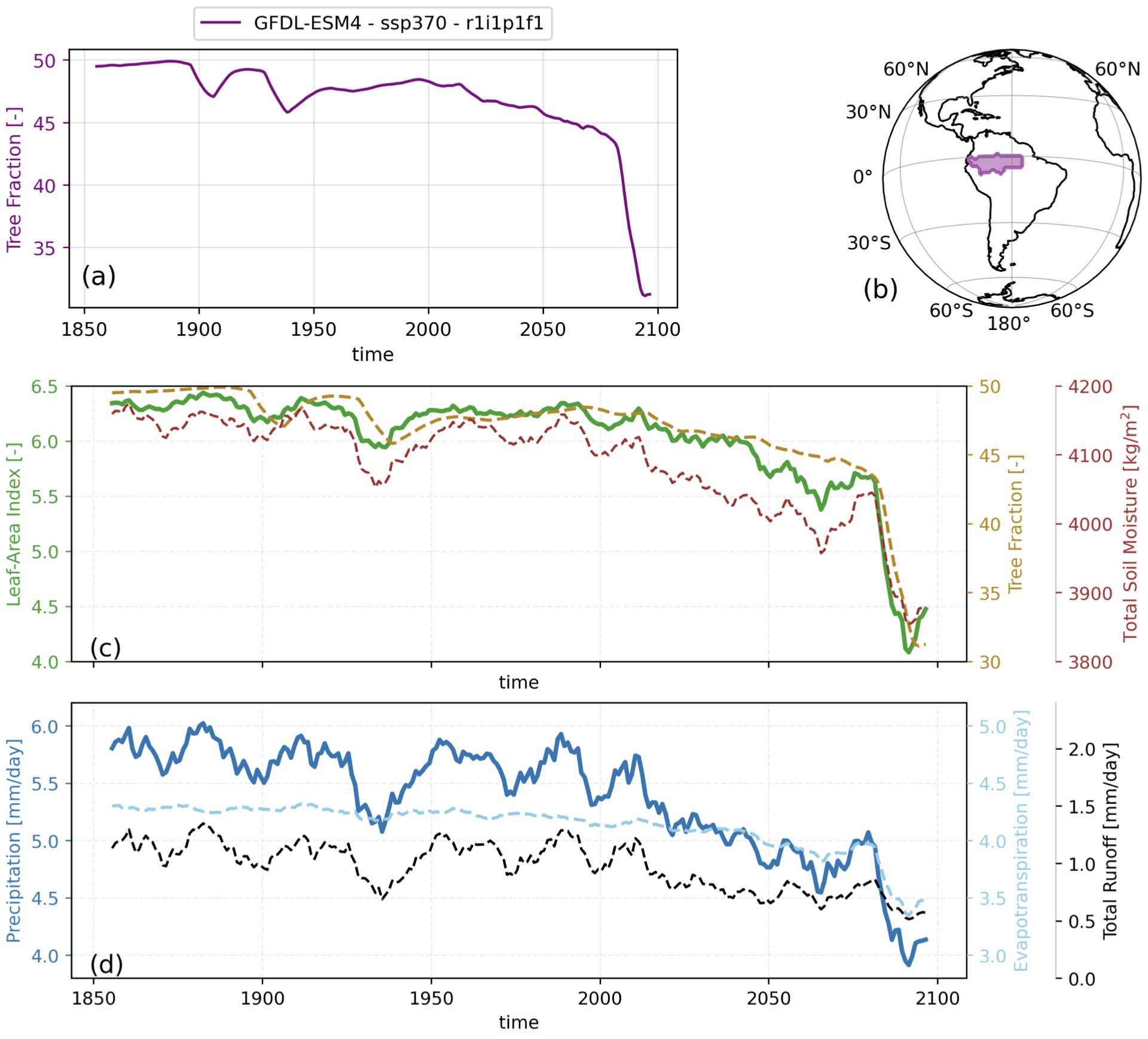}
    \caption{(a) Temporal evolution (10-year running mean) of tree fraction; corresponding spatial extent is shown in (b). (c,d) Time series of land and atmospheric variables, respectively, averaged over the region associated with the abrupt decline in tree fraction in the GFDL-ESM4 model.}
    \label{fig: Amazon - abrupt decline - 2}
\end{figure}

The SSP3-7.0 of the GFDL-ESM4 projects an abrupt decline in tree fraction over the Amazon, albeit over a smaller area compared with the previous sections (Fig. \ref{fig: Amazon - abrupt decline - 2}a,b), accompanied by a consistent reduction in leaf-area index (Fig. \ref{fig: Amazon - abrupt decline - 2}c). Both vegetation parameters respond to a preceding abrupt reduction in soil moisture (Fig. \ref{fig: Amazon - abrupt decline - 2}c). The soil moisture budget attributes this decline to a reduction in precipitation ($\sim$1.1 mm/day), while the compensating declines in evapotranspiration ($\sim$0.6 mm/day) and total runoff ($\sim$0.3 mm/day) are smaller and less abrupt (Fig. \ref{fig: Amazon - abrupt decline - 2}d), leaving a net soil moisture deficit. Horizontal moisture transport diagnostics are unavailable for this model; however, the residual of the atmospheric moisture budget — assuming steady state in total atmospheric moisture — indicates a reduction in moisture convergence concurrent with the precipitation decline.

Interestingly, the SSP5-8.5 scenario of the same model does not qualify as an SNS. Although precipitation declines by $\sim$0.8 mm/day driving a $\sim$200 kg/m$^2$ reduction in soil moisture — comparable in magnitude to SSP3-7.0 but less abrupt — the tree fraction decline is smaller (Fig. \ref{fig: Amazon - abrupt decline - 2 - ssp585}a,b).
% A possible explanation is that the higher CO$_2$ concentrations in SSP5-8.5 enhance vegetation productivity through CO$_2$ fertilization effects, partially offsetting the hydrological stress and preventing tree fraction from crossing the dieback threshold.
The absence of an abrupt decline in SSP5-8.5 despite a comparable precipitation reduction suggests that the stronger CO$_2$ fertilization effect under higher emissions, through increased water use efficiency \citep[e.g.,][]{dekker2016wateruse}, partially offsets hydrological stress, raising the threshold for vegetation tipping. SSP2-4.5, by contrast, shows neither the sharp precipitation decline nor the highly elevated CO$_2$ concentrations of the higher scenarios, and accordingly shows limited vegetation change. Taken together, these three scenarios suggest that the abrupt vegetation decline in SSP3-7.0 may emerge from a specific combination of sufficiently abrupt hydrological stress and insufficient CO$_2$ fertilization to counteract it.

\subsubsection{(A$+$) State transition of the Amazon basin into a higher-biomass state}
\label{sec: amazon greening}

\begin{figure}[t!]
    \centering
    \includegraphics[width=0.7\linewidth]{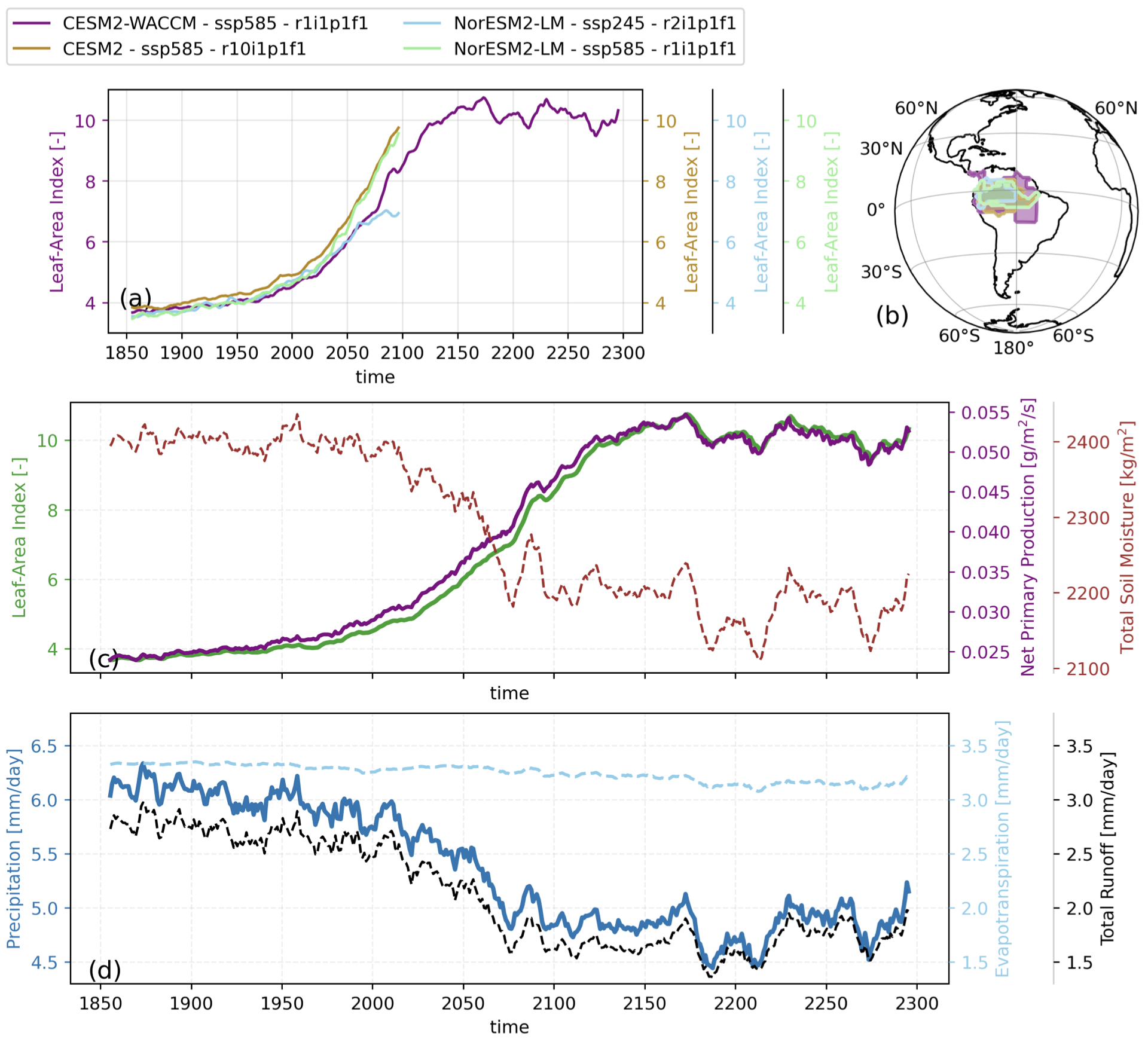}
    \caption{(a) Temporal evolution (10-year running mean) of leaf-area index for different models; corresponding spatial extent is shown in (b). (c,d) Time series of land and atmospheric variables, averaged over the region associated with the state transition (increase) in leaf-area index in the CESM2-WACCM model (purple line in a).}
    \label{fig: Amazon - increase - 2}
\end{figure}

Climate models CESM2 (SSP5-8.5), CESM2-WACCM (SSP5-8.5), and NorESM2-LM (SSP2-4.5, SSP5-8.5), all of which feature the same land-surface parametrization scheme \cite[CLM5, ][]{lawrence2019} which is a non-dynamic vegetation scheme, exhibit SNS behavior over the Amazon basin, where the leaf-area index increases (Fig. \ref{fig: Amazon - increase - 2}a,b), in contrast to the cases discussed in the previous sections. To better understand this behavior, we focus on the CESM2-WACCM simulation, which extends until 2300; the upcoming explanations hold for the CESM2 and NorESM2-LM models (not shown). In CESM2-WACCM, leaf area index closely follows the evolution of net primary production (Fig. \ref{fig: Amazon - increase - 2}c). Soil moisture decreases from around 2000 by approximately 8\% till 2100 (200 kg/m$^2$), yet this reduction does not appear sufficient to induce water stress in vegetation. 

To understand what drives the soil moisture reduction, we examine the soil moisture water budget, in which precipitation is balanced by evapotranspiration, total runoff, and changes in soil moisture storage. During 2000-2100, while the evapotranspiration remains stable, the decline in precipitation is largely absorbed by a reduction in total runoff, with the residual driving the gradual reduction in soil moisture (Fig. \ref{fig: Amazon - increase - 2}c,d). The stability of evapotranspiration, despite the decline in precipitation and soil moisture (Fig. \ref{fig: Amazon - increase - 2}d), reflects the absence of hydrological stress on vegetation: as CO$_2$ concentrations rise under SSP5-8.5, stomatal closure reduces transpiration per unit of leaf area \citep[e.g.,][]{skinner2017role,kennedy2019plantCLM5}, offsetting the increased transpiring surface associated with growing leaf area index and keeping total evapotranspiration approximately constant. As a result, CO$_2$ fertilization remains the dominant driver of vegetation growth, and the system follows a greening trajectory rather than dieback. This obviously contrasts sharply with the dieback models of the previous section, which is further discussed in Section~\ref{sec: amazon inter-model difference}.
% This contrasts sharply with the UKESM1-0-LL model, where a drier initial state in soil moisture means that a comparable soil moisture reduction crosses the stress threshold, suppressing both photosynthesis and transpiration and triggering the precipitation-evapotranspiration-vegetation feedback described above. The difference between the dieback and greening models is further discussed in Section~\ref{sec: amazon inter-model difference}.

\subsubsection{Inter-model differences in Amazon hydrological and vegetation responses}
\label{sec: amazon inter-model difference}

Table \ref{tab:amazon_budget} summarizes the changes in precipitation, evapotranspiration, and soil moisture across the six models and scenarios considered here, together with their relative changes between the first and last 30 years of each simulation. We first compare the five models exhibiting a more gradually developing response (UKESM1-0-LL, IPSL-CM6A-LR, CESM2-WACCM, CESM2, and NorESM2-LM), before turning to the one model with an abrupt response (GFDL-ESM4). Because the land surface schemes underlying these six models differ substantially in prescribed soil column depth -- 2~m in ORCHIDEE \cite[IPSL-CM6A-LR,][]{krinner2005}, $\sim$3~m in JULES \cite[UKESM1-0-LL,][]{best2011}, up to 8.5~m in CLM5 \cite[CESM2, CESM2-WACCM, NorESM2-LM,][]{lawrence2019}, and 10~m in LM4.1 \cite[GFDL-ESM4,][]{shevliakova2024} -- absolute total-column soil moisture (mrso) is not directly comparable across models. We therefore also include the near-surface soil moisture (mrsos), integrated over a common uppermost 10~cm regardless of each model's native soil column, alongside its relative change, as a depth-robust complement to the total-column diagnostics.

Among the more gradually developing state transition cases, UKESM1-0-LL is the clearest example of a straightforward, precipitation-driven response. Its precipitation decline is by far the largest in among these models, both in absolute (-2.94~mm/day) and relative (-48\%, nearly double the next-largest decline, CESM2-WACCM's -26\%) terms. This strong decline propagates through the system: evapotranspiration falls by -1.86~mm/day (-46\%), and soil moisture declines by -28\% at the full-column level and by -66\% at the near-surface, both the largest relative reductions among these models. UKESM1-0-LL's dieback is therefore consistent with a simple picture in which a strong precipitation decline drives correspondingly large reductions in evapotranspiration and soil moisture. Because both the precipitation decline and the moisture response are simultaneously the most extreme of the six models, however, this case alone cannot distinguish whether dieback follows from the magnitude of the reduced precipitation or from the resulting soil moisture deficit specifically -- a distinction the next comparison addresses directly.

IPSL-CM6A-LR provides a more informative comparison, because its precipitation forcing is not the largest of these six models. Its relative decline (-17\%) is smaller than that of CESM2-WACCM (-26\%) and essentially identical to that of CESM2 (-16\%) and NorESM2-LM (-17\%) -- the three models that, under comparable or larger precipitation loss, exhibit greening rather than dieback. Despite this, IPSL-CM6A-LR loses substantially more of its soil moisture than any of the three: -14\% of its full soil column, exceeding even CESM2-WACCM's -12\%, and -22\% of its near-surface layer, well above the entire greening range of -7\% to -13\%. Its evapotranspiration decline (-15\%) likewise exceeds the greening range (-1\% to -12\%). The same precipitation decline that leaves three other models greening under CO$_2$ fertilization instead drives substantially greater relative soil moisture loss in IPSL-CM6A-LR. This suggests that this model crosses a vegetation moisture-stress threshold that the three greening models do not reach, after which evapotranspiration itself declines, reinforcing the deficit and tipping the system toward biomass loss rather than CO$_2$-driven growth. Why a comparable precipitation decline translates into a substantially larger soil moisture deficit in IPSL-CM6A-LR than in the three greening models -- a contrast that persists at the near-surface level, where soil column depth does not confound the comparison -- remains an open question; it plausibly reflects differences in how each land surface scheme partitions precipitation among infiltration, runoff, drainage, and evapotranspiration, but identifying the specific mechanism would require a direct comparison of each model's land-surface formulation, which lies beyond the scope of this multi-model catalogue.

At the total-column level, the GFDL-ESM4 model with an abrupt response appears the most resilient in the ensemble: the largest initial soil moisture (4090~kg/m$^2$) and the smallest relative depletion (-4\%) of all six models. At the near-surface level, this apparent resilience disappears: its initial moisture (29~kg/m$^2$) is indistinguishable from the other five models, and its relative depletion (-15\%) sits just above the greening range (-7\% to -13\%), while its evapotranspiration decline (-16\%) falls within the range spanned by the two dieback cases (-15\% to -46\%). Despite this, and despite a relative precipitation decline (-25\%) comparable to that of CESM2-WACCM (-26\%), a model that greens, GFDL-ESM4's response is abrupt: a relatively fast precipitation decline of similar magnitude to the greening models produces a rapid reduction in soil moisture and vegetation biomass, rather than the more gradually developing transition (decline) seen in UKESM1-0-LL and IPSL-CM6A-LR. This behaviour is consistent with rate-induced tipping, in which the rate of forcing -- rather than its magnitude, or the resulting soil moisture or evapotranspiration state -- exceeds the system's recovery capacity, allowing the system to cross a tipping threshold it would not reach under slower forcing of comparable magnitude. CMIP6 output alone does not allow a formal distinction between rate-induced tipping and alternative mechanisms -- including the possibility, explained below, that both cases cross a common threshold and differ mainly in how rapidly the resulting dieback is realized -- so this interpretation
should be regarded as a hypothesis.

Furthermore, the contrast between more gradually developing transitions and abrupt dieback may partly reflect differences in the representation of the processes involved rather than whether a tipping threshold is crossed. UKESM1-0-LL and GFDL-ESM4, for instance, are the only two models here with a dynamic global vegetation scheme -- TRIFFID \citep{sellar2019ukesm} and LM4.1 \citep{shevliakova2024} -- so that dieback can manifest as an actual loss of tree cover rather than a reduction in biomass on a prescribed distribution. Notably, these two are the only dynamic-vegetation models in the ensemble of models showing an SNS and both dieback, whereas greening occurs only among the climate models with prescribed vegetation. One process that differs even between UKESM1-0-LL and GFDL-ESM4 is fire: LM4.1 represents fire-driven disturbance, a fast feedback capable of rapid, self-reinforcing forest loss, whereas the CMIP6 configuration of UKESM1-0-LL does not. It is therefore plausible that this feedback contributes to GFDL-ESM4's abrupt response, and its absence -- together with the inertia of the vegetation component itself -- to UKESM1-0-LL's more gradually developing one. Following the distinction between committed and realized ecosystem change \citep{jones2009committed}, both models may have crossed a tipping point while differing in how quickly that commitment is realized, a rate set by the processes each model represents. This committed--realized view is an alternative to the rate-induced tipping hypothesis mentioned above: where that hypothesis has GFDL-ESM4 crossing a threshold that slower forcing would not reach, the committed--realized reading has both models crossing the threshold and differing mainly in how quickly the resulting dieback is realized. CMIP6 output alone does not alow to distinguish the two, but under the committed--realized reading the more gradually developing and abrupt responses differ less in their underlying tipping behaviour than their labels suggest.

\begin{table}[t!]
\centering
\tiny
\resizebox{\textwidth}{!}{%
\begin{tabular}{llllrrrrrrrrrr}
\hline
\multirow{2}{*}{Model} & \multirow{2}{*}{Scenario} & \multirow{2}{*}{SNS Type} & \multirow{2}{*}{Response} &
\multicolumn{2}{c}{Precipitation} & \multicolumn{2}{c}{Evapotranspiration} &
\multicolumn{3}{c}{Total Soil Moisture} & \multicolumn{3}{c}{Near-surface Soil Moisture} \\
\cline{5-6} \cline{7-8} \cline{9-11} \cline{12-14}
& & & & $\Delta$ [mm/day] & Rel. $\Delta$ [-] & $\Delta$ [mm/day] & Rel. $\Delta$ [-] &
$\Delta$ [kg/m$^2$] & Initial [kg/m$^2$] & Rel. $\Delta$ [-] &
$\Delta$ [kg/m$^2$] & Initial [kg/m$^2$] & Rel. $\Delta$ [-] \\
\hline
% \rowcolor{purple!15}
UKESM1-0-LL  & SSP5-8.5 & State transition & Dieback  & -2.94 & -0.48 & -1.86 & -0.46 & -266.13 & 961.65  & -0.28 & -20.82 & 31.34 & -0.66 \\
% \rowcolor{purple!15}
IPSL-CM6A-LR & SSP5-8.5 & State transition & Dieback  & -0.89 & -0.17 & -0.56 & -0.15 &  -86.59 & 612.83  & -0.14 &  -6.05 & 27.23 & -0.22 \\
% \rowcolor{green!15}
CESM2-WACCM  & SSP5-8.5 & State transition & Greening & -1.39 & -0.26 & -0.39 & -0.12 & -286.03 & 2417.64 & -0.12 &  -4.06 & 31.74 & -0.13 \\
% \rowcolor{green!15}
CESM2        & SSP5-8.5 & State transition & Greening & -0.85 & -0.16 & -0.03 & -0.01 & -180.38 & 2445.63 & -0.07 &  -2.41 & 31.98 & -0.08 \\
% \rowcolor{green!15}
NorESM2-LM   & SSP5-8.5 & State transition & Greening & -0.84 & -0.17 & -0.20 & -0.06 & -173.06 & 2591.59 & -0.07 &  -2.22 & 31.71 & -0.07 \\
\hline
\hline
% \rowcolor{purple!15}
GFDL-ESM4    & SSP3-7.0 & Abrupt Change  & Dieback  & -1.36 & -0.25 & -0.62 & -0.16 & -162.84 & 4090.14 & -0.04 &  -4.48 & 29.42 & -0.15 \\
\hline
\end{tabular}
}
\caption{Summary of Amazon moisture budget changes across six CMIP6 models under their respective scenarios. Values are spatially averaged over the same purple region shown for the UKESM1-0-LL case in Fig.~\ref{fig: Amazon - decline - 1}a,b, with $\Delta$ and relative changes computed between the first and last 30 years of each simulation. Total soil moisture (mrso) integrates each model's full prescribed soil column (2--10~m, depending on the land model), whereas near-surface soil moisture (mrsos) is integrated over a common uppermost 10~cm and is therefore directly comparable across models.}
\label{tab:amazon_budget}
\end{table}

\subsection{Africa}

% {\color{red}UKESM + MRI + non-dynamic vegetation section + CNRM-mrso for Africa (forgotten!!)}

\subsubsection{(B) State transition of African regions into a higher-biomass state}

\begin{itemize}
    \item \textbf{Transition of Eastern Africa into a higher-vegetation state}
\end{itemize}

\begin{figure}[t!]
    \centering
    \includegraphics[width=0.7\linewidth]{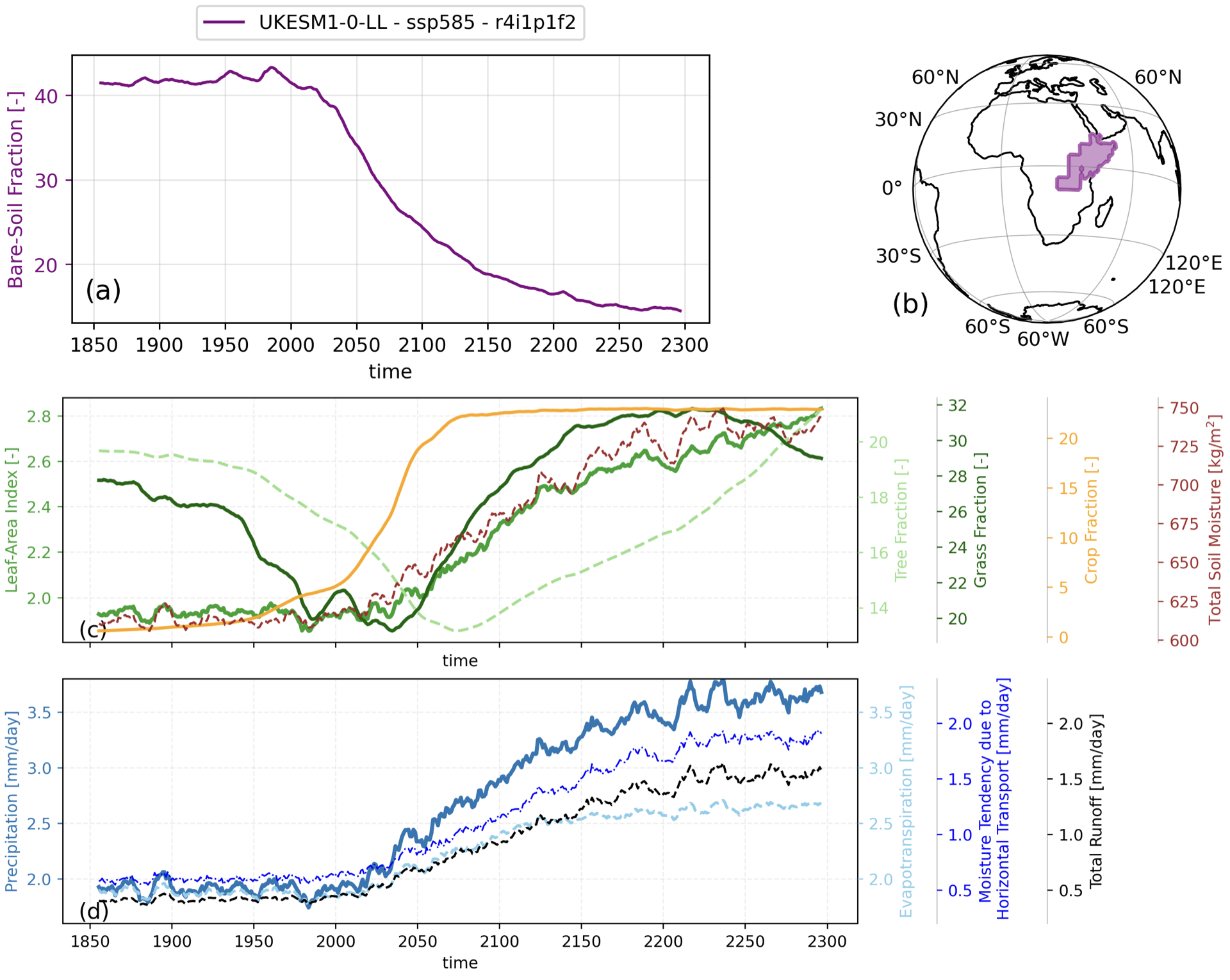}
    \caption{(a) Temporal evolution (10-year running mean) of bare-soil fraction; corresponding spatial extent is shown in (b). (c,d) Time series of land and atmospheric variables, averaged over the region associated with the state transition (decrease) in bare-soil fraction in the UKESM1-0-LL.}
    \label{fig: Africa - decrease - 1}
\end{figure}

\label{sec: africa greening - 1}
The SSP5-8.5 scenario of the UKESM1-0-LL model projects a state transition (decrease) in bare-soil fraction over central and eastern Africa, from around 40\% during 1850–2060 to approximately 10\% by 2300 (Figs. \ref{fig: Africa - decrease - 1}a,b). This change is primarily driven by an increase in crop-area fraction of about 20\% from 1850 to 2100, and partly by increases in tree (5–7\%) and grass (10\%) fractions from 2060 to 2300 (Figs. \ref{fig: Africa - decrease - 1}c). Although crop expansion is driven by the land-use change scenario in the southern part of the region, part of the decrease in bare-soil fraction is due to increases in tree and grass fractions in the northern part (not shown). 

The increase in grass fraction is synchronized with the increase in leaf-area index, which closely follows the increase in soil moisture (Figs. \ref{fig: Africa - decrease - 1}c). Soil moisture, in turn, closely follows the increase in precipitation over the region. Moisture budget analysis of the atmosphere shows that during 1850--2020, precipitation and evapotranspiration remain stable and closely track each other (Figs. \ref{fig: Africa - decrease - 1}d). After 2020, precipitation increases, initially due to enhanced horizontal moisture transport into the region. The increase in precipitation is partitioned into an increase in total runoff and an increase in evapotranspiration, with the rest contributing to an increase in soil moisture, amounting to an increase of roughly 150 kg/m$^2$ in soil water storage (Figs. \ref{fig: Africa - decrease - 1}c,d). This increase in soil moisture drives higher evapotranspiration, which feeds back to further enhance precipitation. Note that after 2020, the combined moisture sources — evapotranspiration and horizontal moisture convergence — exceed precipitation, implying a positive tendency in atmospheric water vapor storage over this region, consistent with a moistening (from 25 kg/m$^2$ in the historical period to 65 kg/m$^2$ by 2250) of the atmospheric column (not shown).

To further explore whether this response is due to the natural feedback of the climate system to climate change, rather than to land-use forcing, we analyzed the same data over this region for the abrupt-4xCO2 scenario (but only for the available member r1i1p1f2 in Figs. \ref{fig: africa baresoil 1 time - abrupt-4xCO2 scenario} and \ref{fig: africa baresoil 1 space - abrupt-4xCO2 scenario}). These figures show that, over the same region where the workflow identified a decrease in bare-soil fraction in the SSP5-8.5 scenario, the abrupt-4xCO2 scenario also exhibits a decline of around 12\% in bare-soil fraction. This decline is driven by a 10\% increase in tree fraction and a 2\% increase in grass fraction. Note that the magnitude of the changes in precipitation and evapotranspiration is weaker than in the SSP5-8.5 scenario, because the forcing is not as strong in this case. This is consistent with the fact that, over this region, the increase in near-surface air temperature is around half of that in SSP5-8.5 (not shown). Despite this difference, these figures suggest that we can consider this case as an SNS, because roughly half of the change in bare-soil fraction in the SSP5-8.5 scenario is due to tree and grass expansion, which consistently occurs in the abrupt-4xCO2 scenario where no land-use forcing is imposed.

\begin{itemize}
    \item \textbf{State transition of the Congo basin into a higher leaf-area-index state}
\end{itemize}
The SSP5-8.5 scenario of the models CanESM5-CanOE, CanESM5-1, CanESM5, and NorESM2-LM projects a transition of the Congo basin into a high-leaf-area-index regime. To understand the underlying processes driving this change, we analyze CanESM5-CanOE as an example. The increase in leaf-area index is consistent with increased net primary production. However, this increase does not appear to be driven through a hydrological pathway. Despite a small increase in precipitation over the region from around 2020, the additional precipitation is partitioned into total runoff rather than soil moisture storage, such that soil moisture shows no consistent trend. Consistently, evapotranspiration remains stable despite the precipitation increase, indicating that vegetation is not experiencing enhanced water availability. The stability of evapotranspiration alongside increasing leaf-area index further suggests that CO$_2$-driven stomatal closure is offsetting the increase in transpiring leaf surface \citep[e.g.,][]{skinner2017role,kennedy2019plantCLM5}, consistent with the behavior identified in the Amazon greening cases. We therefore attribute the increase in leaf-area index primarily to CO$_2$ fertilization effects, which drive enhanced photosynthesis and productivity independently of hydrological conditions. The explanation above also applies to CanESM5-1, CanESM5, and NorESM2-LM, which show a response similar to that of CanESM5-CanOE (not shown).

\begin{figure}[t!]
    \centering
    \includegraphics[width=0.7\linewidth]{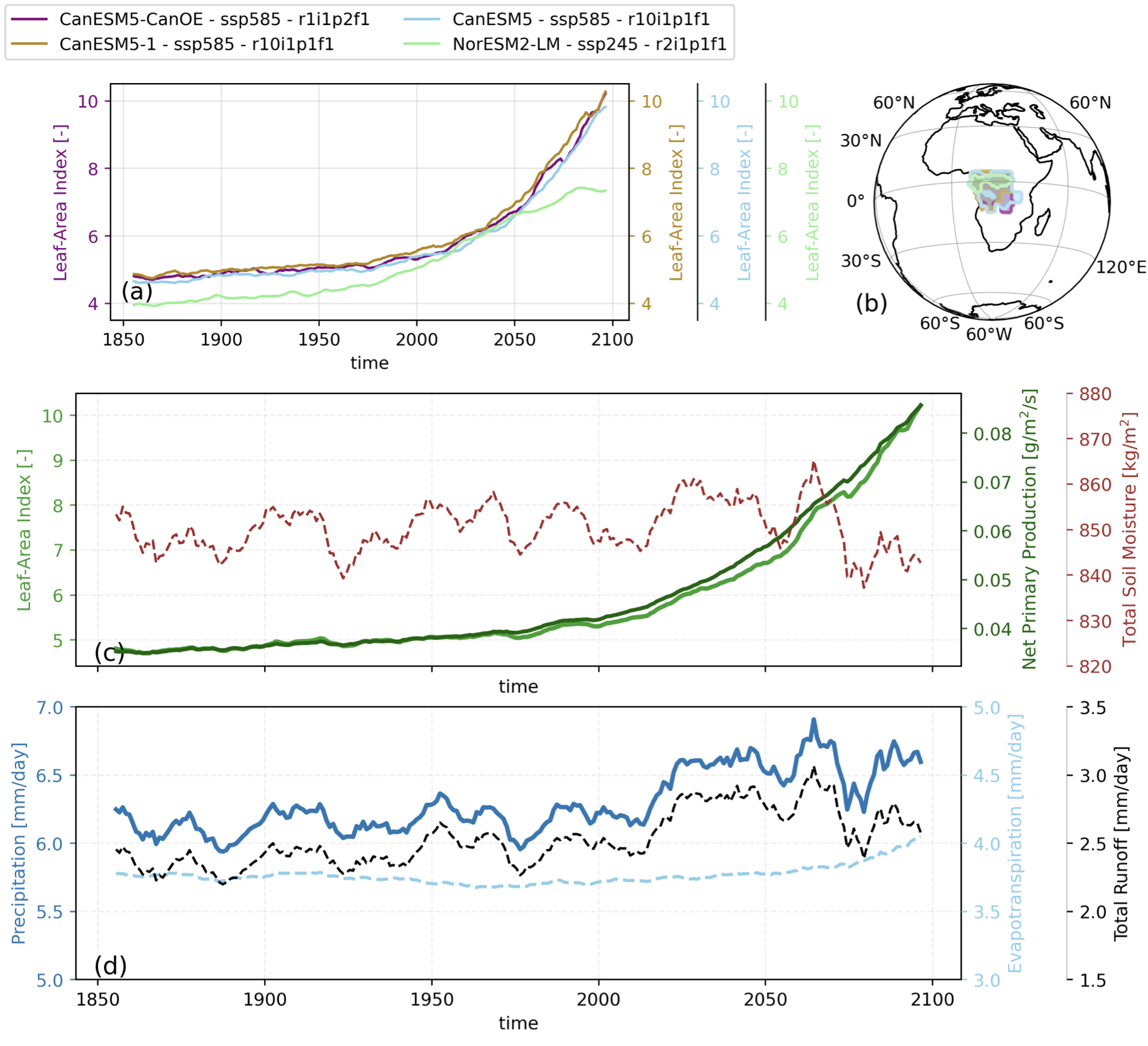}
    \caption{(a) Temporal evolution (10-year running mean) of leaf-area index for couple of models; corresponding spatial extent is shown in (b). (c,d) Time series of land and atmospheric variables, averaged over the region associated with the state transition (increase) in leaf-area index in the CanESM5-CanOE.}
    \label{fig: Africa - increase - 3}
\end{figure}

\subsubsection{(b) Abrupt soil-moisture decline in the Congo basin}
\label{sec: africa drying}

\begin{figure}[t!]
    \centering
    \includegraphics[width=0.7\linewidth]{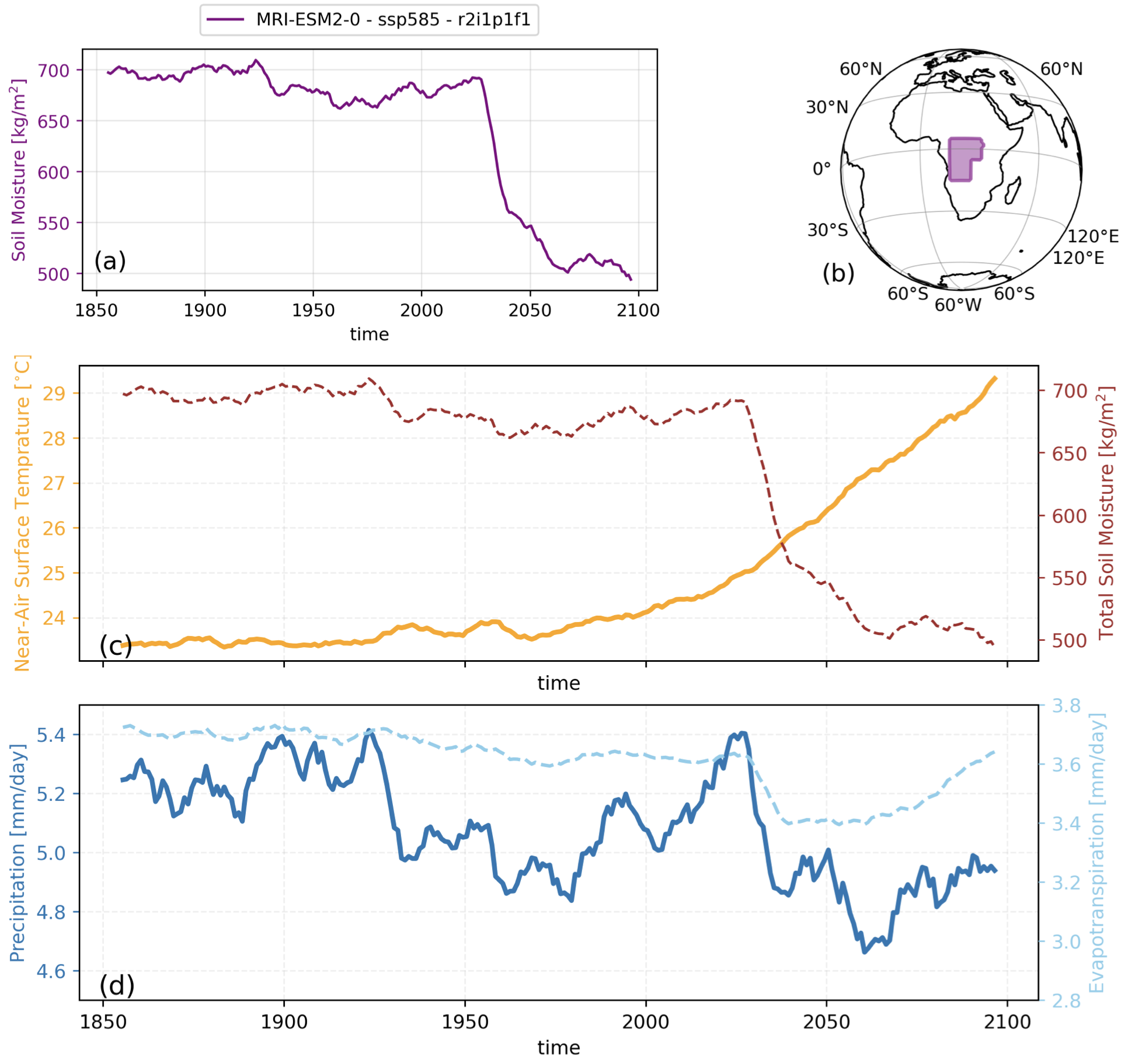}
    \caption{(a) Temporal evolution (10-year running mean) of soil moisture; corresponding spatial extent is shown in (b). (c,d) Time series of land and atmospheric variables, averaged over the region associated with the abrupt decrease in soil moisture in the MRI-ESM2-0.}
    \label{fig: Africa - decrease - 2}
\end{figure}

The SSP5-8.5 scenario of the MRI-ESM2-0 model (r2i1p1f1) shows an abrupt decrease in soil moisture over the Congo basin, from 700 to 500~kg~m$^{-2}$
around 2040 (Fig. \ref{fig: Africa - decrease - 2}a,b). The first member (r1i1p1f1) exhibits a similarly sharp decline (not shown); however, it does not satisfy all abruptness criteria, specifically because its maximum jump is not 5 or 10 times larger than the standard deviation of the tropical domain in the combined historical–scenario or piControl time series.

In member r2i1p1f1 that is categorized as an SNS, despite considerable decadal variability, both precipitation and evapotranspiration exhibit reductions around the same period around 2030 (Fig. \ref{fig: Africa - decrease - 2}c,d), and spatial maps comparing the first and last thirty years confirm that these declines co-occur with that of soil moisture in the same areas (not shown). However, because the MRI-ESM2-0 model does not output the vertically integrated zonal and meridional moisture fluxes required for the atmospheric moisture budget decomposition, we cannot determine whether the precipitation decline is driven by reduced horizontal moisture transport into the region or by reduced local evapotranspiration. Consequently, two interpretations remain plausible: the precipitation decline may be the primary driver of the soil moisture reduction, with evapotranspiration responding subsequently; alternatively, warming-induced increases in evaporative demand may have depleted soil moisture first, with declines in evapotranspiration and precipitation following as a consequence. Time series of precipitation and evapotranspiration (Fig. \ref{fig: Africa - decrease - 2}d) show that changes in evapotranspiration occur much more slowly than those in precipitation and soil moisture, suggesting that the first mechanism is more plausible. Distinguishing between these mechanisms would require either moisture flux outputs or targeted sensitivity experiments, which are beyond the scope of the present analysis.

\subsection{Arctic zone}
\subsubsection{(C) Transition of the boreal forest into an expanded-vegetation state}
\label{sec: boreal forest}

\begin{figure}[t!]
    \centering
    \includegraphics[width=0.7\linewidth]{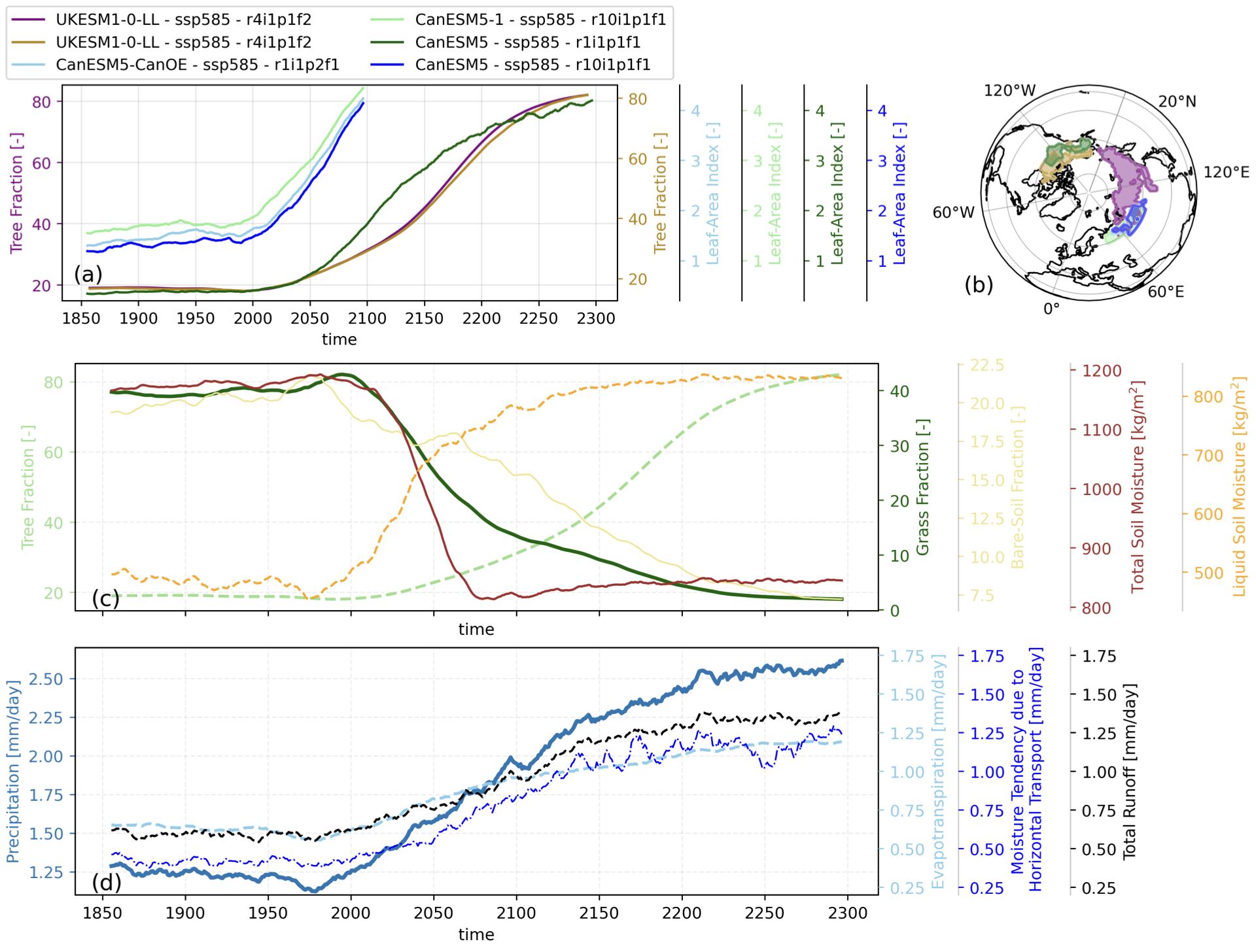}
    \caption{(a) Temporal evolution (10-year running mean) of vegetation-related parameters for couple of models; corresponding spatial extent is shown in (b). (c,d) Time series of land and atmospheric variables, averaged over the region associated with the state transition (increase) in tree fraction in the UKESM1-0-LL.}
    \label{fig: Arctic - increase - 1}
\end{figure}

Models UKESM1-0-LL (SSP5-8.5, dynamic vegetation) and a family of CanESM5 model variants (SSP5-8.5, non-dynamic vegetation) are identified by our workflow as projecting a strong nonlinear increase in tree fraction and leaf-area index across the high-latitude regions of northern North America and Asia (Fig. \ref{fig: Arctic - increase - 1}a,b). To understand the drivers, we analyze the UKESM1-0-LL model over northern Asia as an example. This case shows a significant increase in tree fraction from approximately 20\% during 1850--2050 to 80\% by 2300 (purple case in Fig. \ref{fig: Arctic - increase - 1}a,b), accompanied by the complete disappearance of grass-area fraction (by 40\%) and a 20\% reduction in bare-soil fraction to nearly zero (Fig. \ref{fig: Arctic - increase - 1}c). The expansion of trees is synchronous with the abrupt disappearance of frozen soil-water content due to warming (not shown). Although total soil moisture decreases (Fig. \ref{fig: Arctic - increase - 1}c), permafrost thaw means that water in the uppermost soil layers transitions from the frozen to the liquid phase (Fig. \ref{fig: Arctic - increase - 1}c), creating favorable conditions for root water uptake and tree growth. The warming-driven thaw therefore alleviates the primary limitation on tree expansion in this region.

Atmospheric moisture budget analysis (Fig. \ref{fig: Arctic - increase - 1}d) shows that precipitation increases simultaneously with tree expansion. However, this precipitation increase is most likely not the cause of tree expansion but rather a consequence of it: horizontal moisture transport into the region remains relatively stable during the initial expansion phase, while evapotranspiration rises in tandem with the growing tree fraction, driving the concurrent increase in precipitation through enhanced local moisture recycling. At a later stage, horizontal moisture transport also increases, further amplifying precipitation and helping to sustain the favorable conditions for continued tree expansion. This behavior is consistent with previous studies showing widespread high-latitude greening and tree expansion under warming scenarios \citep{pugh2018carbon}. The resulting increase in tree cover may further amplify regional warming through reductions in surface albedo \citep{falloon2012role}, representing a positive feedback between vegetation change and warming over the region.

This mechanism also appears to be consistent across the CanESM5 family of models. Although CanESM5 uses a non-dynamic vegetation parameterization (unlike UKESM), it exhibits a similar response: both precipitation and evapotranspiration increase consistently with increasing leaf-area index (not shown) over the corresponding regions shown in Fig.~\ref{fig: Arctic - increase - 1}b. However, frozen soil moisture is not available for these models, preventing a direct assessment of the role of permafrost thaw in this model family here.

\subsubsection{(D,d) State transition of land permafrost into a thawed state}
\label{sec: permafrost}
\begin{figure}[t!]
    \centering
    \includegraphics[width=0.7\linewidth]{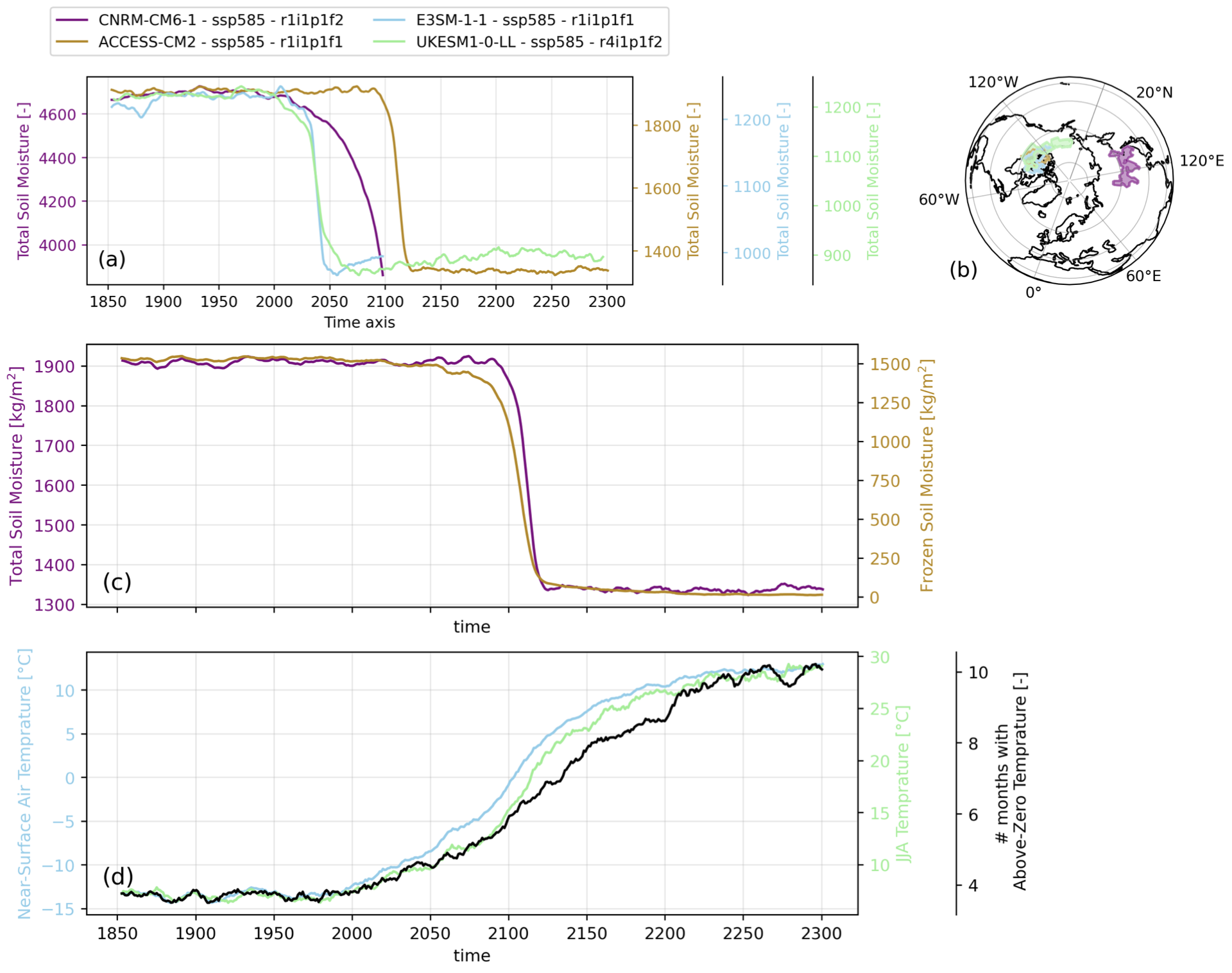}
    \caption{(a) Temporal evolution (10-year running mean) of total soil moisture for couple of models; corresponding spatial extent is shown in (b). (c,d) Time series of land and atmospheric variables, averaged over the region associated with the abrupt decrease in soil moisture in the ACCESS-CM2.}
    \label{fig: Arctic - decrease - 2}
\end{figure}

The climate model CNRM (variants CNRM-CM6-1, CNRM-CM6-1-HR, and CNRM-ESM2-1) projects a strong decline in total soil water content over the high latitudes of northern Asia and North America. While all three variants exhibit this transition under SSP5-8.5, the variants CNRM-CM6-1 and CNRM-CM6-1-HR display it consistently across all other scenarios as well, including SSP1-2.6, SSP2-4.5, and SSP3-7.0. Similar but more abrupt transitions in total soil moisture over the same regions are projected by the climate models ACCESS (variants ACCESS-ESM1-5 and ACCESS-CM2), E3SM-1-1, and UKESM1-0-LL (SSP3-7.0 and SSP5-8.5). Representative time series and their associated regions are shown in Fig. \ref{fig: Arctic - decrease - 2}a,b.

In all the models listed above, the decrease in total soil moisture is primarily driven by a strong reduction in frozen soil water content, which is approximately twice the magnitude of the total soil moisture decrease (example for the ACCESS-CM2 model in Fig. \ref{fig: Arctic - decrease - 2}c). This implies that roughly half of the water released by thawing remains in the soil in liquid form, while the remainder drains out. The reduction in frozen soil water is driven by rising temperatures and generally occurs when (i) the yearly-mean near-surface air temperature approaches 0$^{\circ}$C and (ii) the number of months per year with monthly-mean temperatures above zero exceeds approximately five (Fig. \ref{fig: Arctic - decrease - 2}d). In cases with abrupt transitions, the frozen water content completely disappears within a few decades of the onset of thaw. We note that near-surface air temperature serves here as a proxy for the more physically direct quantity, soil or ground surface temperature, which governs permafrost thaw through the crossing of the melting point at depth. 

An important point concerns the climate models TaiESM1 and NorESM2 (variants NorESM2-LM and NorESM2-MM), which project a strong nonlinear \textit{increase} in total soil water content over the same region. Closer inspection reveals that these models include only the liquid phase in their total soil moisture (mrso) output, excluding frozen soil water (mrfso) entirely. When the frozen water content is analyzed separately (Fig. \ref{fig: arctic mrso 2 time}), a decrease is indeed found, consistent with the other models. However, unlike the cases described above, the meltwater does not drain from the soil column but instead remains stored in the liquid phase (mrlso), resulting in the apparent increase in mrso. This analysis shows that the output convention regarding soil moisture must be accounted for when interpreting total soil moisture changes in high-latitude regions.

\subsubsection{(E) State transition of far northeastern America into a low snow-cover state}
\label{sec: snow decline}
\begin{figure}[t!]
    \centering
    \includegraphics[width=0.7\linewidth]{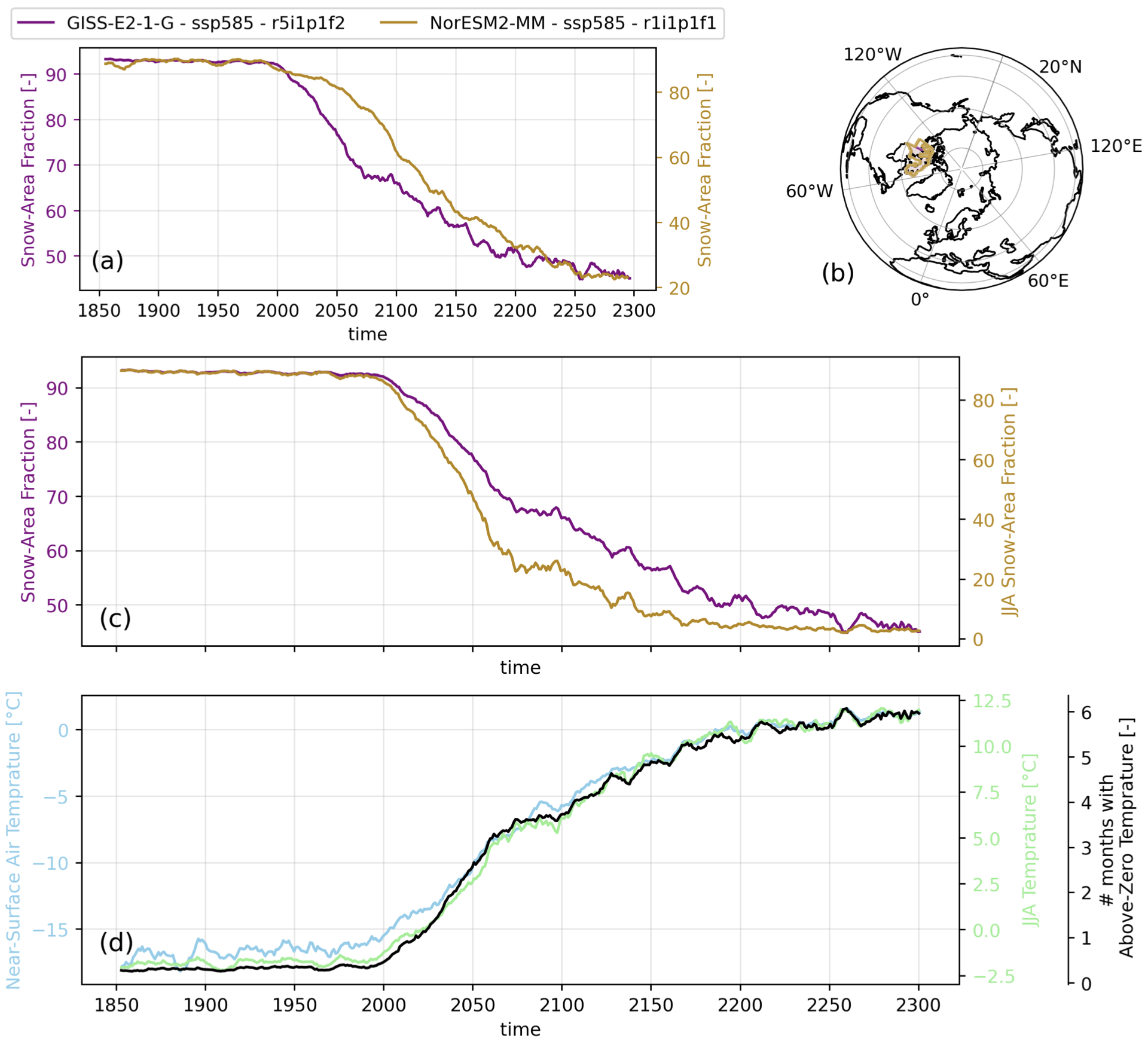}
    \caption{(a) Temporal evolution (10-year running mean) of snow-area fraction for couple of models; corresponding spatial extent is shown in (b). (c,d) Time series of land and atmospheric variables, averaged over the region associated with the abrupt decrease in soil moisture in the GISS-E2-1-G.}
    \label{fig: Arctic - decrease - 3}
\end{figure}

The SSP5-8.5 scenario of models GISS-E2-1-G and NorESM2-MM projects a strong nonlinear decrease in snow-area fraction, from near-full coverage over the corresponding region during 1850--2000 to approximately 45\% (GISS-E2-1-G) and 20\% (NorESM2-MM) in the yearly mean over the far northeast of North America (Fig. \ref{fig: Arctic - decrease - 3}). Figure \ref{fig: Arctic - decrease - 3}c,d shows time series for the GISS-E2-1-G model, illustrating that the decrease begins after the summer-mean (June, July, August – JJA) near-surface air temperature crosses 0°C, coinciding with the first occurrence of monthly-mean temperatures exceeding 0°C within a year. The summer-mean (JJA) snow-area fraction completely disappears by 2200 in the region. The loss of snow cover is likely to further amplify regional warming through the snow-albedo feedback, whereby reduced surface reflectivity increases the absorption of incoming solar radiation, accelerating local warming and reinforcing further snow loss \citep[e.g.,][]{pithan2014arctic}. The NorESM2-MM model exhibits a similar temperature response, with the nonlinear decline in snow-area fraction likewise beginning after summer-mean (JJA) near-surface air temperature exceeds 0°C (not shown).

\subsection{Asia and the northeast Northern America}

\subsubsection{(F) State transition in biomass over Asia}
\label{sec: asia greening}
\begin{figure}[t!]
    \centering
    \includegraphics[width=0.7\linewidth]{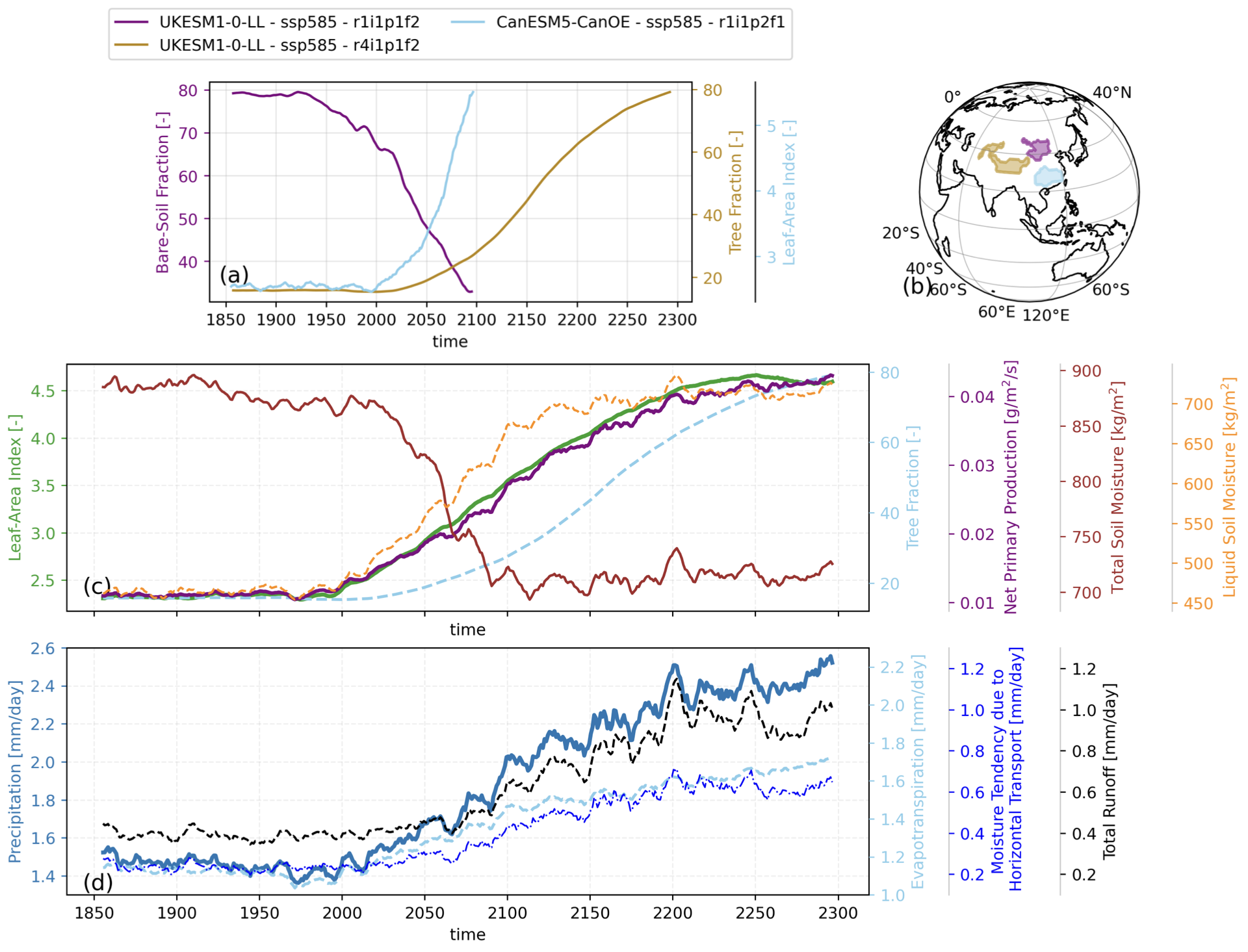}
    \caption{(a) Temporal evolution (10-year running mean) of land-vegetation-related variables for couple of models; corresponding spatial extent is shown in (b). (c,d) Time series of land and atmospheric variables, averaged over the region associated with the state transition (increase) in tree fraction in the UKESM1-0-LL model.}
    \label{fig: AsiaUS - increase - 1}
\end{figure}

Several climate models project strong nonlinear vegetation growth and a strong nonlinear reduction in bare-soil fraction over parts of Asia: UKESM1-0-LL (dynamic vegetation; SSP5-8.5), multiple variants of CanESM5 (non-dynamic vegetation; variants: CanESM5, CanESM5-1, CanESM5-CanOE; SSP5-8.5), NorESM2-LM (non-dynamic vegetation; SSP5-8.5), and variants of CESM2 (non-dynamic vegetation; variants: CESM2, CESM2-WACCM; SSP5-8.5). Representative time series and associated regions are shown in Fig. \ref{fig: AsiaUS - increase - 1}a,b.

The brown case in Fig. \ref{fig: AsiaUS - increase - 1}a,b shows tree fraction increasing from approximately 10\% in the historical period to near 80\% by 2300 over a region close to and including the Tibetan Plateau. Here, increases in leaf-area index and net primary production precede the expansion of tree fraction (Fig. \ref{fig: AsiaUS - increase - 1}c). This vegetation growth is concurrent with increased liquid water availability in the soil (Fig. \ref{fig: AsiaUS - increase - 1}c): from around 2000, declining frozen soil moisture drives a reduction in total soil moisture (not shown), but because the total soil moisture decline is approximately half that of the frozen water loss, the liquid water content of the soil must be increasing. Warming therefore creates favorable conditions for vegetation growth by increasing liquid water availability in the root uptake zone. The atmospheric moisture budget further shows that precipitation increases only after evapotranspiration rises, with limited contribution from horizontal moisture transport until around 2050, after which the transport term also drives increased precipitation (Fig. \ref{fig: AsiaUS - increase - 1}d). Total runoff absorbs most of the increase in precipitation, leaving soil moisture stable after 2100. Together, these results indicate that vegetation growth in this region is initiated by warming alleviating soil water limitation rather than by changes in precipitation.

The purple case in Fig. \ref{fig: AsiaUS - increase - 1}a,b shows a strong nonlinear decline in bare-soil fraction driven by the expansion of grass-area fraction (not shown). The decline in this case starts early, already in the historical period, and accelerates under SSP5-8.5, while crop and pasture fractions remain unchanged (not shown). Because vegetation cover is prognostic in UKESM1-0-LL (emerging from dynamic-vegetation competition), this shift occurs despite negligible local trends in temperature and precipitation, and most likely reflects CO\textsubscript{2} fertilization and improved water-use efficiency allowing grass to colonize bare ground in this water-limited dryland.

For the non-dynamic vegetation models projecting increased leaf-area index over Asia, the mechanism is different: over this lower latitude region (compared with the other cases above) soil moisture decreases by around 60 kg/m$^2$, while the frozen soil water content only decreases by 2 kg/m$^2$, implying that warming does not meaningfully increase water availability in the root zone over this region (Fig. \ref{fig: China - gradual increase - 2}). While direct temperature effects on plant growth cannot be fully excluded, the combination of worsening water availability and strong CO$_2$-driven increases in water use efficiency \citep[e.g.,][]{dekker2016wateruse} points to CO$_2$ fertilization as the dominant driver of the observed leaf-area index increase.

\subsubsection{(G) Transition into a higher leaf-area-index state over the northeast Northern America}
\label{sec: usa greening}
\begin{figure}[t!]
    \centering
    \includegraphics[width=0.7\linewidth]{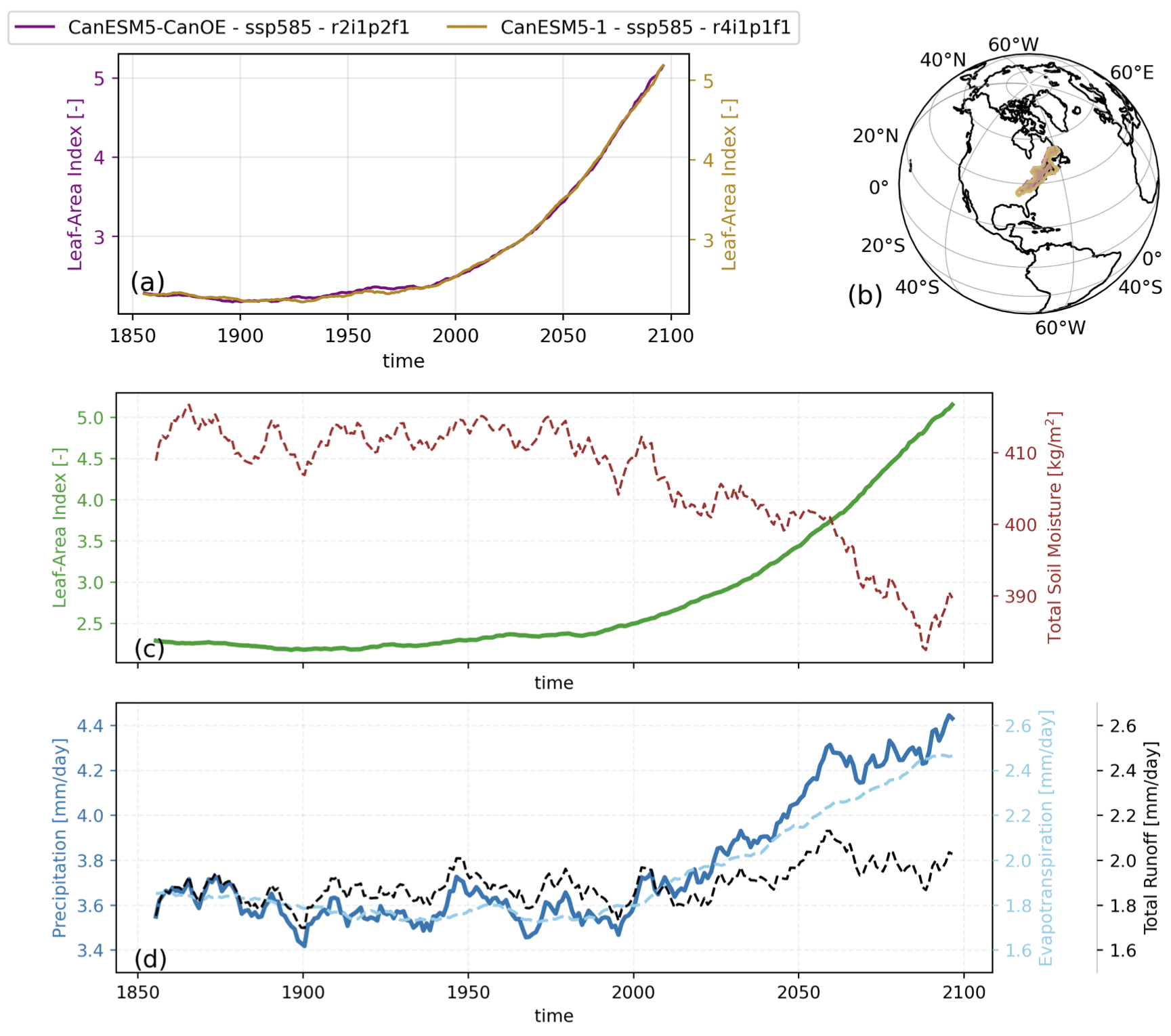}
    \caption{(a) Temporal evolution (10-year running mean) of leaf-area index for a family of CanESM5 models; corresponding spatial extent is shown in (b). (c,d) Time series of land and atmospheric variables, averaged over the region associated with the state transition (increase) in leaf-area index in the CanESM5-CanOE model.}
    \label{fig: AsiaUS - increase - 2}
\end{figure}

The SSP5-8.5 scenario of CanESM5 model variants (CanESM5-1 and CanESM5-CanOE) projects an increased leaf-area index over the northeast Northern America (Fig. \ref{fig: AsiaUS - increase - 2}a,b). Analyzing CanESM5-CanOE as an example, the leaf-area index does not follow the evolution of soil moisture, which begins to decline around 1980, while the leaf-area index continues to grow almost exponentially from 1920 onward (Fig. \ref{fig: AsiaUS - increase - 2}c). The evolution of the land moisture budget terms shows that the increase in evapotranspiration is synchronous with that of the leaf-area index (Fig. \ref{fig: AsiaUS - increase - 2}d). This synchronization along with the decline in the soil moisture suggest that the increase in precipitation is most likely a response to enhanced evapotranspiration driven by vegetation growth. However, this interpretation would be more robust if the moisture tendency due to horizontal transport convergence were available, which is not the case for this model. Total runoff remains relatively stable, but its slight increase, together with the increased evapotranspiration, outweighs the precipitation flux, resulting in a small negative soil moisture storage tendency (Fig. \ref{fig: AsiaUS - increase - 2}c,d). In essence, this case indicates that the vegetation response in this region and model is not driven by hydrological changes; rather, it is primarily driven by CO$_2$ fertilization effects, which are strongest under the SSP5-8.5 scenario. Note that increased CO$_2$ concentrations drive stomatal closure, reducing transpiration per unit leaf area \citep[e.g.,][]{skinner2017role,kennedy2019plantCLM5}; however, the concurrent increase in leaf-area index drives greater evapotranspiration overall, and this effect outweighs the stomatal closure-induced decline in evapotranspiration, resulting in a net increase in region-averaged evapotranspiration.

\subsection{Warming levels at which SNSs occur and reach maximum change}
\label{sec:warming_levels}

\begin{figure}[H]
    \centering
    \includegraphics[width=0.8\linewidth]{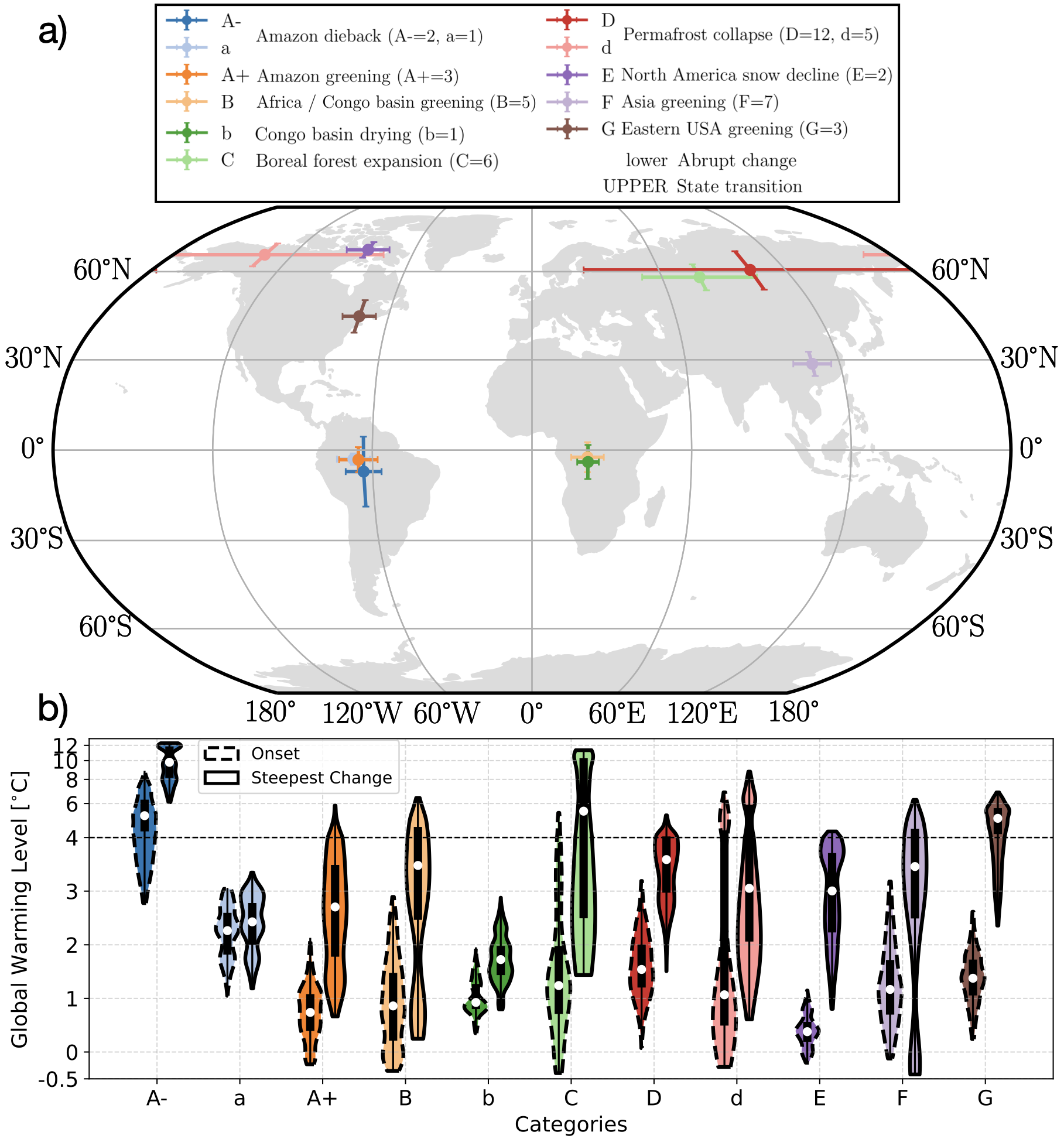}
    \caption{(a) Global map of all Strong Nonlinear Surprises (SNSs) identified in this study. Uppercase and lowercase letters denote state transitions and abrupt changes, respectively. The numbers in parentheses indicate the total number of SNS cases (model--region combinations, Table \ref{tab: all SNS cases} reports variables and scenarios associated with each case, and Table \ref{tab: fraction of members giving SNS} shows the fraction of members showing the SNS per model, scenario and variable for each case) within each category. Marker positions correspond to the weighted mean center of all flagged grid cells within a category, with error bars indicating one weighted standard deviation in latitude and longitude. (b) Global warming level relative to 1850–1880 at which each SNS category exhibits its onset (dashed) and steepest rate of change (continuous). Distributions are derived from all contributing window sizes, model members, scenarios, and variables. The white circle denotes the median; the thick black bar spans the interquartile range (25th–75th percentile). Thin black lines extend to the most extreme values within 1.5 times the interquartile range beyond the 25th and 75th percentiles (Tukey whiskers).}
    \label{fig: land - overall}
\end{figure}

With the individual SNS categories described in the preceding sections, we turn to a synthesis of their geographic distribution (Fig.~\ref{fig: land - overall}a) and the warming levels at which each undergoes its most rapid transition as well as levels at which each SNS begins to unfold (Fig.~\ref{fig: land - overall}b). It is important to note that the warming levels corresponding to the moment of steepest change are higher than those of the onset of the transition; the latter already begins at lower warming levels in each case (Fig.~\ref{fig: land - overall}b). Per-case, scenario, member and variable distributions are provided in the appendix (Fig.~\ref{fig: warming levels per case}, and Table~\ref{tab: gwl per member}).

The 50\% inter-percentile ranges of the warming levels span a wide range across categories (Fig.~\ref{fig: land - overall}b); we report these below as onset / steepest-change pairs: 4.2--6.1 / 8.0--12$^{\circ}$C for the state transition of the Amazon basin via forest dieback (A$-$); 1.8--2.6 / 2.0--2.8$^{\circ}$C for abrupt Amazon dieback (a); 0.4--1.1 / 1.8--3.5$^{\circ}$C for Amazon transition toward a greener forest (A$+$); 0.2--1.5 / 2.5--4.7$^{\circ}$C for Congo basin greening (B); 0.9--1.2 / 1.4--2.0$^{\circ}$C for abrupt Congo basin drying (b); 0.8--2.5 / 2.5--10.0$^{\circ}$C for boreal forest expansion (C); 1.2--2.0 / 3.0--4.0$^{\circ}$C for transition of the land permafrost into a thawed state (D); 0.5--4.5 / 2.0--6.0$^{\circ}$C for abrupt permafrost collapse (d); 0.2--0.5 / 2.2--3.7$^{\circ}$C for snow-area decline in North America (E); 0.7--1.7 / 2.5--4.5$^{\circ}$C for Asia greening (F); and 1.1--1.7 / 4.2--5.8$^{\circ}$C for northeast Northern America greening (G).

The width of the distributions in Fig.~\ref{fig: land - overall}b reflects contributions from multiple sources of uncertainty. First, the window size used to identify the period of steepest change or the SNS onset -- which ranges from 20 to 50 years -- introduces spread in both the timing and the associated warming level for each individual SNS. Second, within a given model--scenario combination, corresponding SNSs of different ensemble members or variables may reach their steepest rate of change or begin at different warming levels, adding a further layer of spread. Third, different scenarios within the same model can likely shift the timing of the steepest change or the onset, and fourth, given a category, models themselves can disagree substantially on the warming level at which a given transition starts or its change maximizes. Broadly speaking, inter-model differences tend to dominate the overall spread (Fig.~\ref{fig: warming levels per case}). A clear example is category d, where the timing of abrupt permafrost collapse varies considerably across models: the ACCESS model family exhibits a later collapse than either UKESM1-0-LL or E3SM-1-1 (Figs. \ref{fig: Arctic - decrease - 2}a, \ref{fig: warming levels per case}). Additionally, a category in this catalogue can encompass multiple distinct sub-regions that are grouped together based on their overall geographical locations -- for instance, category F combines Asia greening cases spanning the Tibetan Plateau and East Asia (Fig. \ref{fig: AsiaUS - increase - 1}a,b) -- and differences in the local ecosystem response across these sub-regions contribute an additional source of spread to the category-level distributions. Beyond differences in the simulated transition itself, two models with otherwise identical SNS time series can still disagree on the associated warming level simply because they have different equilibrium climate sensitivities --- that is, different global mean surface temperature responses to a doubling of CO$_2$, driven primarily by differences in low-cloud feedbacks \cite[e.g., ][]{bony2005marine,schneider2017climate,zelinka2020cloud}.

With global warming beyond 0$^{\circ}$C, the majority of SNSs begin to unfold, with onset frequency peaking between 1 and 2$^{\circ}$C (Fig.~\ref{fig: land - overall}b); two categories unfold later, however: Amazon transition toward a drier state (A$-$), peaking around 5$^{\circ}$C, and abrupt Amazon dieback (a), peaking around 2.2$^{\circ}$C. Steepest change follows a similar pattern, increasingly occurring from approximately 0.5--1.5$^{\circ}$C of warming and peaking between 2.5--4.0$^{\circ}$C for the majority of categories.

A notable exception is category F, whose steepest-change distribution shows a spurious secondary peak near 0$^{\circ}$C, corresponding to case 44 (Table~\ref{tab: all SNS cases}), in which the bare-soil fraction in UKESM1-0-LL already declines around 1920. For this case, the onset warming level is in fact slightly higher than that of steepest change: as we expect, the onset year indeed precedes the year of steepest change (Fig.~\ref{fig: warming levels per case}) as required mathematically, but this early in the historical record the forced warming trend is weak enough that internal variability can locally reverse the year-to-warming-level ordering (Fig.~\ref{fig: AsiaUS - increase - 1}a,b and Fig.~\ref{fig: warming levels per case}). Category F's primary peak, however, falls within its expected 50\% inter-percentile range, consistent with the remaining cases in the category. Category B similarly exhibits a bimodal onset-and-steepest-change distribution, where the earlier peak is associated with case 10 (Table~\ref{tab: all SNS cases}), in which leaf-area index increases in NorESM2-LM over a substantially smaller region than the other cases in this category (Fig.~\ref{fig: Africa - increase - 3}a,b).

Many of the SNS categories identified here occur under SSP5-8.5 or SSP3-7.0 (Fig.~\ref{fig: warming levels per case}), beginning to unfold already in the last decades of the historical period or the first few decades of the scenario period -- a window during which all five scenarios remain fairly similar. What distinguishes SSP5-8.5 and SSP3-7.0 is not that these transitions start earlier under them, but that only these higher-emission pathways continue warming long enough afterward for the transitions to complete: SSP1-1.9, SSP1-2.6, and SSP2-4.5 rarely reach the warming levels several categories require to reach their steepest change, so these transitions remain incomplete, or undetected as SNS cases, under lower-emission futures. This suggests that the ambition of mitigation, especially how much warming is ultimately avoided, is a powerful lever for reducing the risk of such surprises.

% Finally, the large spread in category A$-$ is driven primarily by the UKESM1-0-LL ensemble, where multiple members differ slightly in the geographical extent of the transition and in the variables that trigger it, resulting in a range of steepest-change timings and consequently a wide warming-level distribution (Fig.~\ref{fig: warming levels per case}).

% \newpage
\section{Summary \& Discussion}
\label{sec:discussion}

This study, using the automatic detection workflow developed by \cite{angevaare2025catalogue} and applied to the CMIP6 ensemble under shared socio-economic pathway emission scenarios, identified 47 SNS cases across the land-vegetation component of the climate system, classified into 9 categories (Table \ref{tab: all SNS cases}). Over the Amazon basin, two cases show a state transition via forest dieback (A$-$, Section \ref{sec: grad amazon dieback}); one case shows an abrupt Amazon dieback (a, Section \ref{sec: abrupt amazon dieback}); and three cases project a state transition into a higher-biomass state (A$+$, Section \ref{sec: amazon greening}). The contrast between the two dieback–greening trajectories is explored in Section \ref{sec: amazon inter-model difference} and traced primarily to differences in precipitation: where precipitation declines similarly across models, dieback occurs in those where the same relative drop in precipitation translates into a larger decline in soil moisture, especially close to the surface (Table \ref{tab:amazon_budget}); otherwise, the CO$_2$ fertilization effect wins out and drives greening. This precipitation decline in dieback-prone models appears to be dynamically driven -- through a weakening of convective mass flux -- since the thermodynamic component alone would, if anything, favor more precipitation given the warming-induced increase in moisture over the Amazon. In addition, models representing dynamic vegetation distributions show only dieback, whereas greening is only limited to models without dynamic vegetation distributions. The third category emerged over Africa (B, Section \ref{sec: africa greening - 1}): two cases of greening, one over eastern-central Africa driven by increased precipitation, and one over the Congo basin driven mainly by CO$_2$ fertilization. The fourth, a contrasting category (b, Section \ref{sec: africa drying}), is an abrupt soil-moisture drying over the Congo basin, possibly linked to a precipitation decline in that model. The fifth category (C, Section \ref{sec: boreal forest}) is the expansion of boreal forest across the high-latitude band, governed primarily by warming, which relieves growth-limiting conditions such as the lack of liquid water in the root-uptake zone. The sixth category covers permafrost collapse at high latitudes, both state transitions (D) and abrupt changes (d), driven by warming and typically triggered once the duration of above-zero temperatures exceeds half the year (Section \ref{sec: permafrost}). The seventh category (E, Section \ref{sec: snow decline}) is a transition to a lower snow-cover state over northeastern North America, driven by warming and accelerated by the regional snow-albedo feedback. The eighth category (F, Section \ref{sec: asia greening}) covers greening near the Tibetan Plateau, where warming improves liquid-water availability for vegetation, and over southeastern Asia via CO$_2$ fertilization. The ninth and final category (G, Section \ref{sec: usa greening}) is a transition to a higher leaf-area index state over the northeast Northern America, driven primarily by CO$_2$ fertilization. Taken together, the interquartile ranges of the steepest-change warming level vary widely across categories, from as low as 1.4$^{\circ}$C to above 12$^{\circ}$C (Fig.~\ref{fig: land - overall}b), showing that the timing of these shifts depends as much on the nature of the underlying process as on the magnitude of warming itself. Onset warming levels are correspondingly lower for every category, spanning from as low as 0.2$^{\circ}$C to 6.1$^{\circ}$C overall (Section~\ref{sec:warming_levels}).

Before discussing the physical mechanisms behind individual categories, it is worth acknowledging the methodological choices that shape the catalogue presented here. The detection framework of \cite{angevaare2025catalogue} involves several numerical decisions that carry inherent trade-offs. The minimum spatial area criterion is one such choice: it is designed to retain physically coherent, region-wide ($>10^6$~km$^2$) signals while filtering out spurious small-scale anomalies — a deliberate design decision meaning that more localized SNSs, such as those identified by \cite{parry2025amazon} over the Amazon, fall below our threshold by construction. Lowering this criterion would recover such local signals but at the cost of robustness against noise. Similarly, binning model output into discrete warming levels assumes that ensemble members provide a meaningful sample of the forced response at each level, an assumption that becomes increasingly uncertain for SNSs detected in only one or two models. Beyond these numerical choices, the physical drivers proposed throughout this study are inferred from diagnostic analysis of model output rather than from targeted process-level experiments; dedicated sensitivity experiments including but not limited to isolating, for instance, the role of soil moisture, land-scheme or convective parameterization depending on the component — would be needed to rigorously test the proposed mechanisms. These caveats are particularly relevant for the categories where inter-model spread is large and the underlying processes are strongly parameterization-dependent, as we discuss below.

The response of tropical rainforest including Amazonian basin to warming is strongly modulated by the coupling between precipitation, soil moisture, and evapotranspiration — a feedback loop that is parameterized, rather than explicitly resolved, in the coarse-resolution models analyzed here. The strength of this coupling varies considerably across the CMIP6 ensemble, which, as discussed in Section \ref{sec: amazon inter-model difference}, is a leading source of inter-model spread in the Amazon projections. This concern is not merely structural: recent evidence from global storm-resolving models with non-parameterized but explicit convection shows that this land-atmosphere coupling is substantially weaker than what coarse models imply \citep{lee2024precip}, suggesting that the strong feedbacks driving dieback in some CMIP6 models may be an artifact of convective parameterization rather than a robust physical signal.

An important point to consider is that the Amazon-related warming levels reported here, clearly suggesting Amazon dieback is unlikely to occur, do not account for deforestation, since we do not treat direct land-use change as a \textit{surprise}. Deforestation can nonetheless drive self-perpetuating surprises beyond the cleared areas: by weakening the evapotranspiration–precipitation feedback discussed above, it reduces moisture recycling along low-level moisture transport trajectory and can thereby dry downwind regions that were not themselves deforested \cite[e.g.,][]{butt2026tropical}. By reducing condensational heating, it also weakens the import of moisture from the Atlantic into the Amazon basin, reinforcing the hydrological pressure \citep[e.g.,][]{boers2017amazon}. Independently, observation-based dynamical-systems analyses suggest the Amazon may be approaching a tipping point in response to deforestation \citep[e.g.,][]{bochow2023amazon,wunderling2025amazon}. Yet, a recent study based on global storm-resolving models, finds that full deforestation drives a strengthening of large-scale moisture convergence that largely offsets the moisture loss associated with reduced evapotranspiration, effectively buffering the system against its own forcing \citep{yoon2025amazon}. That said, such storm-resolving simulations are necessarily short in duration, meaning they cannot capture the slowly evolving, fully coupled climate in which cascading interactions between tipping elements \citep{wunderling2023overshoot} unfold over decades. In that broader context, weakening of the Atlantic meridional overturning circulation (AMOC) has been linked to a southward shift of the inter-tropical convergence zone \citep{guo2026itcz}, which could increase precipitation over the southern Amazon \citep{benyami2024amoc}, and separately to changes in the seasonality of Amazon rainfall \citep{rene2024amoc}. Taken together, these results highlight that a consensus on the Amazon's fate under deforestation and global warming has not yet emerged, and that single-model or single-method perspectives are unlikely to be sufficient to resolve it.

The permafrost collapse cases identified in this study (categories D and d, Section \ref{sec: permafrost}) are among the most consistently reproduced signals across the CMIP6 ensemble, which itself reflects a broader consensus in the literature: the physical mechanism — warming drives soil thaw, which in turn releases stored carbon — is robust and directionally unambiguous, unlike the contested sign of Amazon vegetation change. That said, important uncertainties remain, particularly regarding the magnitude of the carbon-cycle feedback associated with permafrost thaw \cite[e.g.,][]{schuur2015climate}. CMIP6 models differ considerably in their representation of soil thermal dynamics, the insulating role of snow, and — critically — the size of their soil organic carbon stocks, which are generally underestimated relative to observational estimates \cite[e.g.][]{bruke2020perma}. Moreover, most models do not represent abrupt thaw processes such as thermokarst formation, which can release carbon on timescales far shorter than the more gradually developing thaw captured here \cite[e.g.][]{turetsky2020perma}. The permafrost carbon feedback is therefore likely larger and potentially faster than what the CMIP6 ensemble implies, meaning that while the occurrence of permafrost collapse is a robust projection, its full climatic consequences may be systematically underestimated in the models we analyze.

Finally, a few broader limitations and open questions deserve mention. The SNSs catalogued here are identified in strongly forced systems driven by SSP emission scenarios, which means that, strictly speaking, they cannot be equated with tipping points from the dynamical systems perspective: our framework diagnoses abrupt or nonlinear transitions in model output, but says nothing about their reversibility, implying that the onset-associated warming levels cannot be necessarily considered tipping points. Whether these transitions would persist, or partially recover, under stabilized or declining forcing remains an open question that targeted experiments such as those in the TIPMIP protocol \citep{jones2025tipmip} are well placed to address, for instance through ramp-up and ramp-down forcing designs that explore the reversibility of individual transitions. It is also worth noting that several land components known to be important tipping elements are not yet represented in the CMIP6 models analyzed here — most notably the Greenland and Antarctic ice sheets. Their inclusion in future Earth system model ensembles \cite[e.g.,][]{guo2025greenland} would be a valuable extension of this type of catalogue, not least because the Greenland ice sheet is a key mediator of freshwater fluxes into the North Atlantic and therefore directly relevant to cascading interactions with the AMOC.

Despite these limitations, this study makes a contribution that we believe is both scientifically and societally relevant. By applying a systematic SNS detection framework to the full CMIP6 ensemble, we identify a broad range of potential tipping behaviour across the land-vegetation system -- including not only abrupt transitions, which have received most attention in the tipping-points literature, but also more gradually developing nonlinear shifts that complement the abrupt changes emphasized in dynamical-systems-oriented studies \cite[e.g.,][]{bathiany2020tipping,terpstra2025tipping}. Together, and complementing previous assessments \cite[e.g.,][]{mckay2022tipping,terpstra2025tipping,loriani2025tipping,drijfhout2015abrupt}, these results provide a comprehensive picture of the warming levels at which such transitions emerge across models -- a type of information that is directly actionable for policymakers designing mitigation and adaptation strategies. The same picture also indicates where vigilance is most needed: by flagging the regions and variables most prone to abrupt or more gradually developing transitions, the catalogue can help prioritize the development of early-warning indicators of state transitions \cite[e.g.,][]{scheffer2009early, boulton2022pronounced,tao2023amazon} and motivate sustained observational monitoring in these hotspots, so that approaching transitions might be anticipated rather than merely recorded after the fact.

Yet, perhaps the clearest message that emerges from the full body of results, mechanisms, and uncertainties discussed here is a simple one: across every region and every category -- from Amazon dieback to permafrost collapse to boreal forest expansion -- it is the relentless accumulation of CO$_2$ in the atmosphere that is the common driver, pushing interconnected components of the Earth system into states that carry profound and possibly difficult-to-reverse risks for generations to come. If there is one conclusion that transcends all the complexity documented here, it is that mitigation remains our most powerful tool for reducing those risks.

\newpage

\section*{Code and Data Availability}
The CMIP6 data is publicly available at \url{https://esgf-node.llnl.gov/search/cmip6/}. The software associated with the automatic SNS detection algorithm of \cite{angevaare2025catalogue} is publicly available at \url{https://github.com/JoranAngevaare/optim_esm_tools}. The maps were created in Python using the Cartopy \citep{Cartopy} library.

\section*{Author Contributions}
The concept for this study was developed by PA and SD. Downloading the data as well as the initial SNS detection workflow run on the land–vegetation systems were performed by JA. Sensitivity tests were performed by PA. PA analyzed and visualized the data. PA and SD interpreted the results with contributions from CJ, AW, and JA. PA drafted the initial manuscript, and all authors contributed to its revision and finalization.\\

\section*{Acknowledgments}
We thank the World Climate Research Programme's Working Group on Coupled Modelling, which oversees CMIP, and the climate modelling groups (see Table~\ref{tab: all available data}) for generating and sharing their model output. Coordinating support for CMIP, together with the leadership of software infrastructure development, was provided by the U.S. Department of Energy's Program for Climate Model Diagnosis and Intercomparison, in collaboration with the Global Organization for Earth System Science Portals. We are grateful to the modelling groups for making their output available, to the Earth System Grid Federation (ESGF) for archiving and distributing the data, and to the various funding agencies that support CMIP6 and ESGF. Access to and analysis of the CMIP6 data were carried out on the Dutch supercomputer cluster SURFsara.

PA and SD are supported by TipESM, and SD and JA were supported by OptimESM. TipESM ("Exploring Tipping Points and Their Impacts Using Earth System Models") is funded by the European Union under Grant Agreement No. 101137673. OptimESM ("Optimal High Resolution Earth System Models for Exploring Future Climate Change") is funded by the European Union under Grant Agreement No. 101081193. SD also received support from ISOTIPIC ("Interacting ice Sheet and Ocean Tipping – Indicators, Processes, Impacts and Challenges"), funded by the UK National Environment Research Council under Grant No. NE/Y503320/1.

Generative AI tools were used to assist with grammar and spell checking, minor text revisions, and debugging portions of the analysis code.

\newpage
\bibliographystyle{apalike}
\bibliography{sample}

\newpage
\newpage

\setcounter{page}{1}                      % restart page numbering at 1
\renewcommand{\thepage}{\arabic{page}}   % optional: render them as S1, S2, ...
\section*{Supplementary Information}
\setcounter{figure}{0}
\renewcommand{\thefigure}{S\arabic{figure}}
\setcounter{table}{0}
\renewcommand{\thetable}{S\arabic{table}}

\subsection*{Supplementary Figures}    %% Appendix A
% \appendixfigures

\begin{figure}[H]
    \centering
    \includegraphics[width=\linewidth]{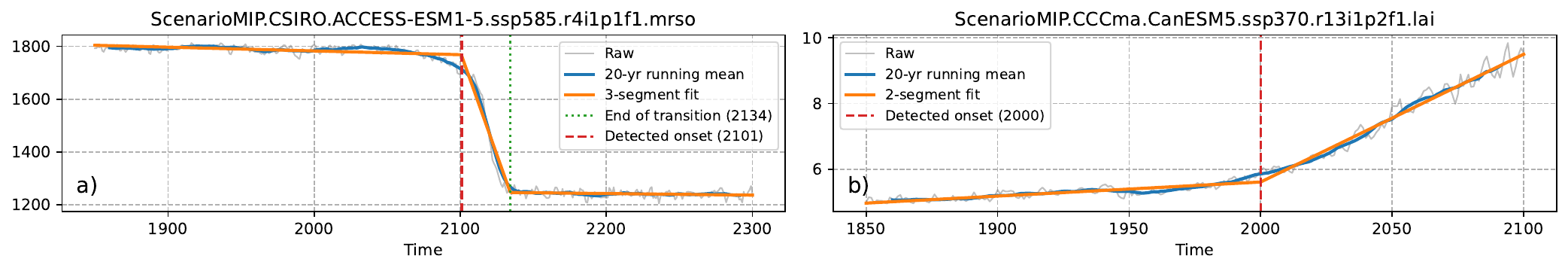}
    \caption{An example of a two-segment and three-segment fit to the SNS time series for the detection of the SNS onset based on a window size of 20 years. See Section~\ref{sec:methods_warming} for the in-depth explanations. For example, for panel a, we assign equal probability to years 2101$\pm$10, since the window size is 20 years here.}
    \label{fig: example fits}
\end{figure}

\begin{figure}[H]
    \centering
    \includegraphics[width=0.7\linewidth]{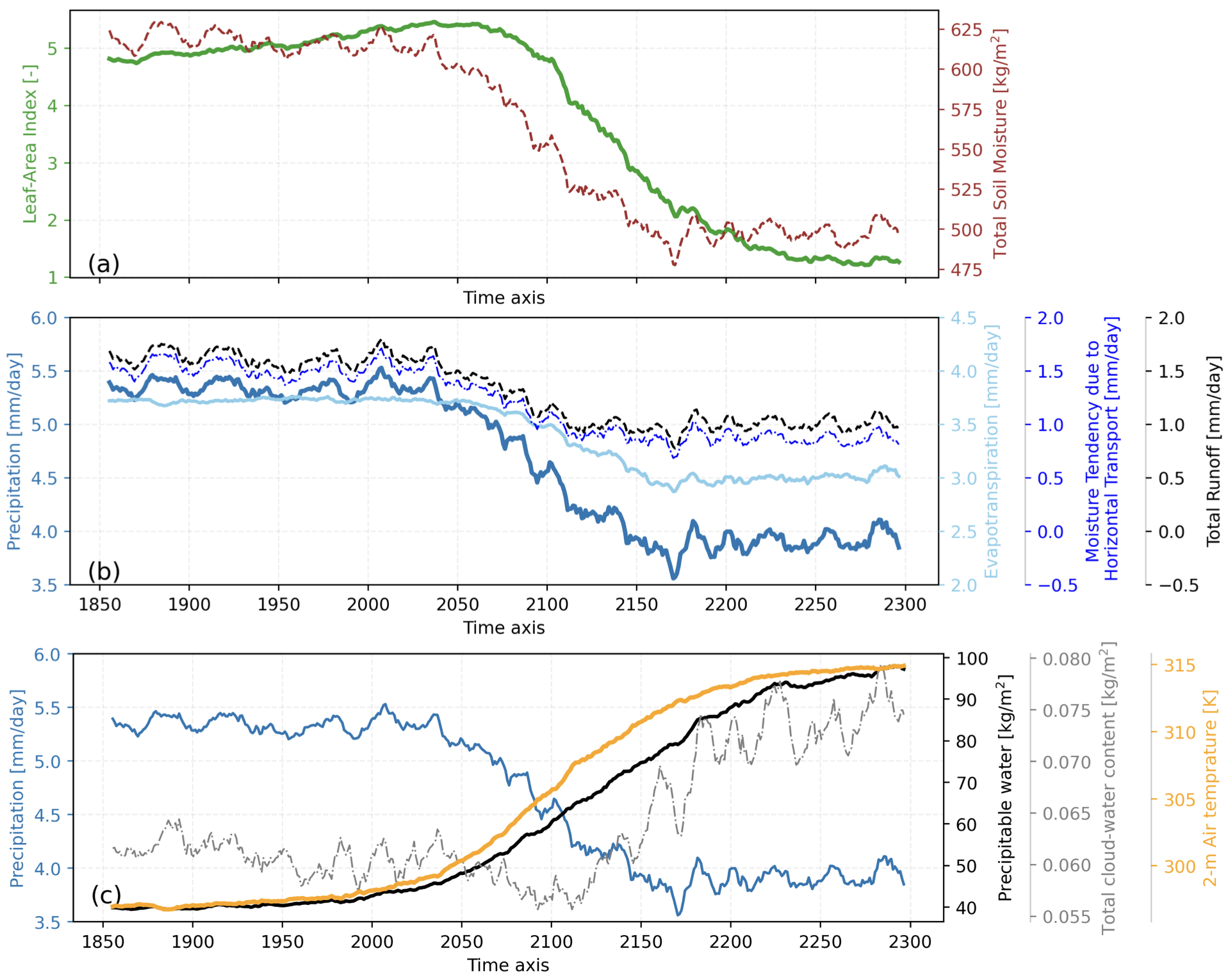}
    \caption{(a-c) Time series of land and atmospheric variables, respectively, averaged over the region associated with the gradual decline in leaf-area index in the IPSL model (green line in Fig. \ref{fig: Amazon - decline - 1}a).}
    \label{fig: Amazon - decline - ipsl}
\end{figure}

\begin{figure}[H]
    \centering
    \includegraphics[width=0.7\linewidth]{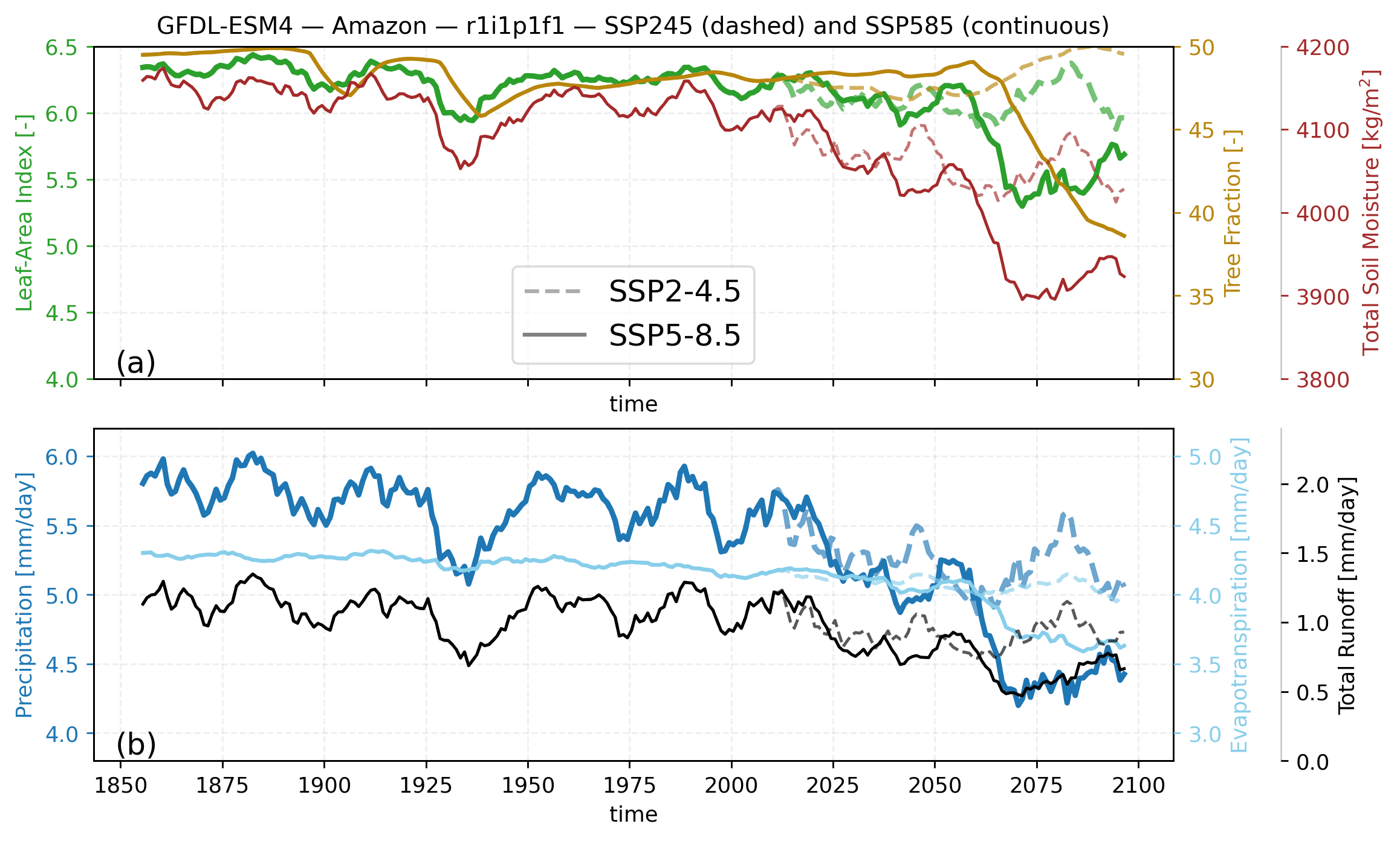}
    \caption{Time series of land and atmospheric (a,b) variables, respectively, averaged over the region associated with the abrupt decline in tree fraction in the GFDL-ESM4 model (SSP3-7.0 in Fig. \ref{fig: Amazon - abrupt decline - 2}) but now for the SSP2-4.5 and SSP5-8.5 scenarios.}
    \label{fig: Amazon - abrupt decline - 2 - ssp585}
\end{figure}

\begin{figure}[H]
    \centering
    \includegraphics[width=\linewidth]{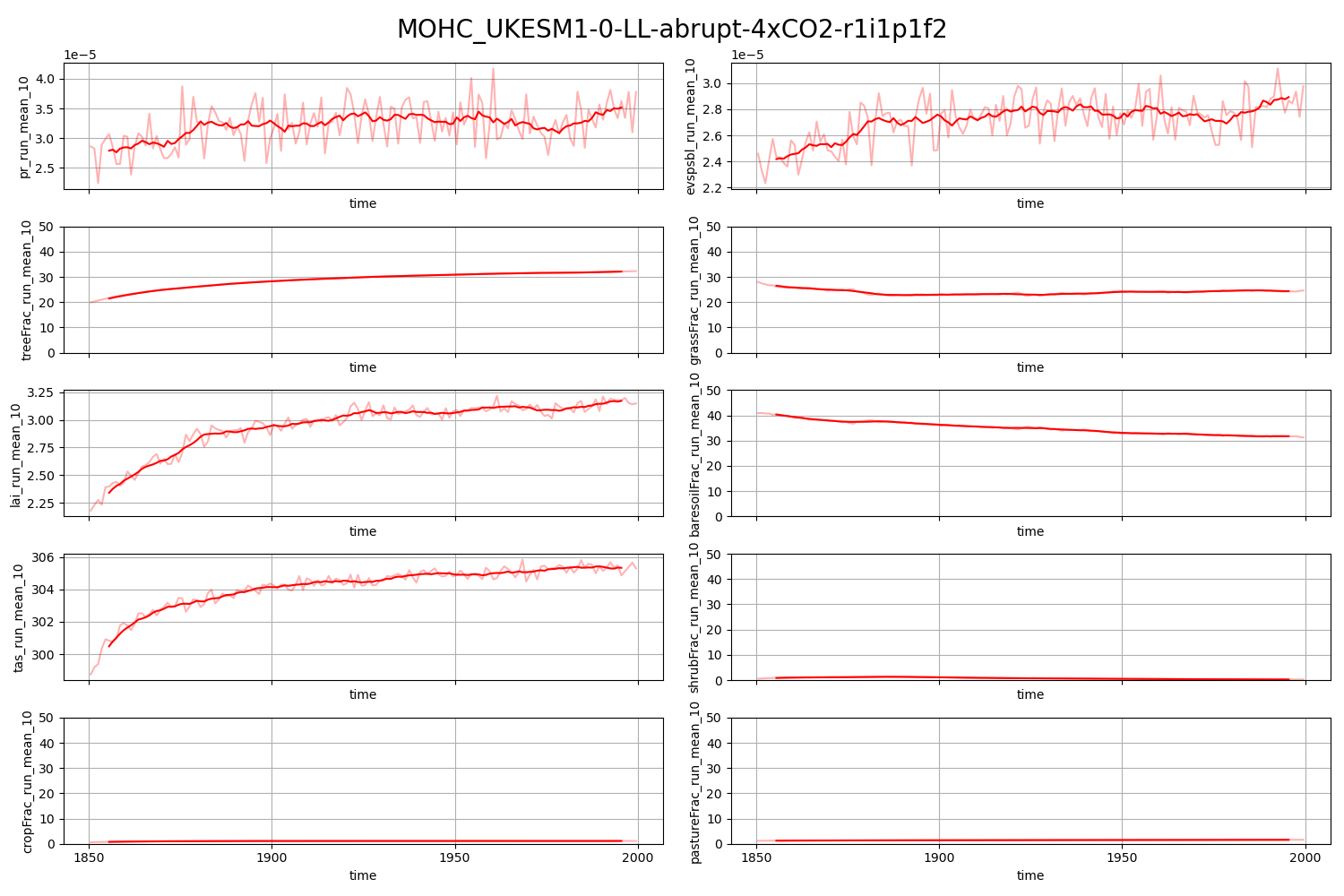}
    \caption{Time series of the case: gradual decrease in bare-soil fraction abrupt-4xCO2 scenario. The corresponding time series are all averaged over the region shown in Figs. \ref{fig: Africa - decrease - 1}b and \ref{fig: africa baresoil 1 space - abrupt-4xCO2 scenario}.}
    \label{fig: africa baresoil 1 time - abrupt-4xCO2 scenario}
\end{figure}

\begin{figure}[H]
    \centering
    \includegraphics[width=\linewidth]{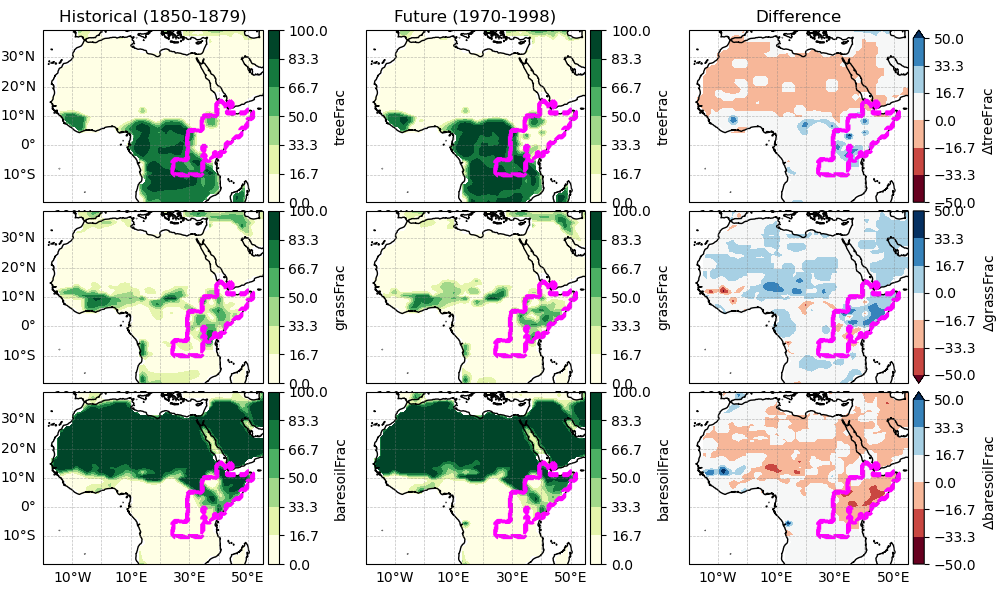}
    \caption{Spatial maps of the case: gradual decrease in bare-soil fraction in abrupt-4xCO2 scenario of the UKESM1-0-LL model.}
    \label{fig: africa baresoil 1 space - abrupt-4xCO2 scenario}
\end{figure}

\begin{figure}[H]
    \centering
    \includegraphics[width=\linewidth]{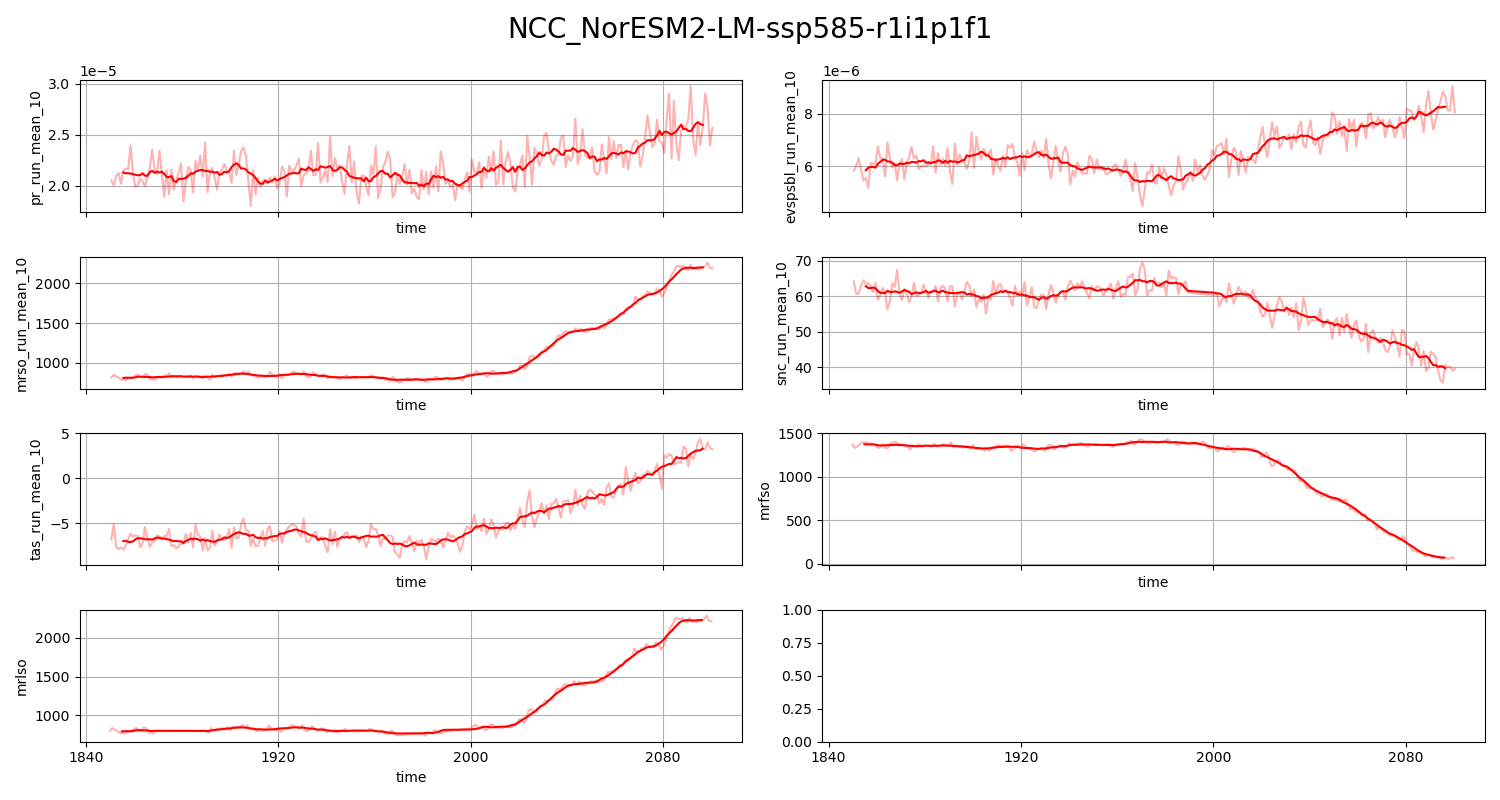}
    \caption{Time series of an example model for the case: gradual increase in total soil water content over Arctic zone}
    \label{fig: arctic mrso 2 time}
\end{figure}

\begin{figure}[H]
    \centering
    \includegraphics[width=0.7\linewidth]{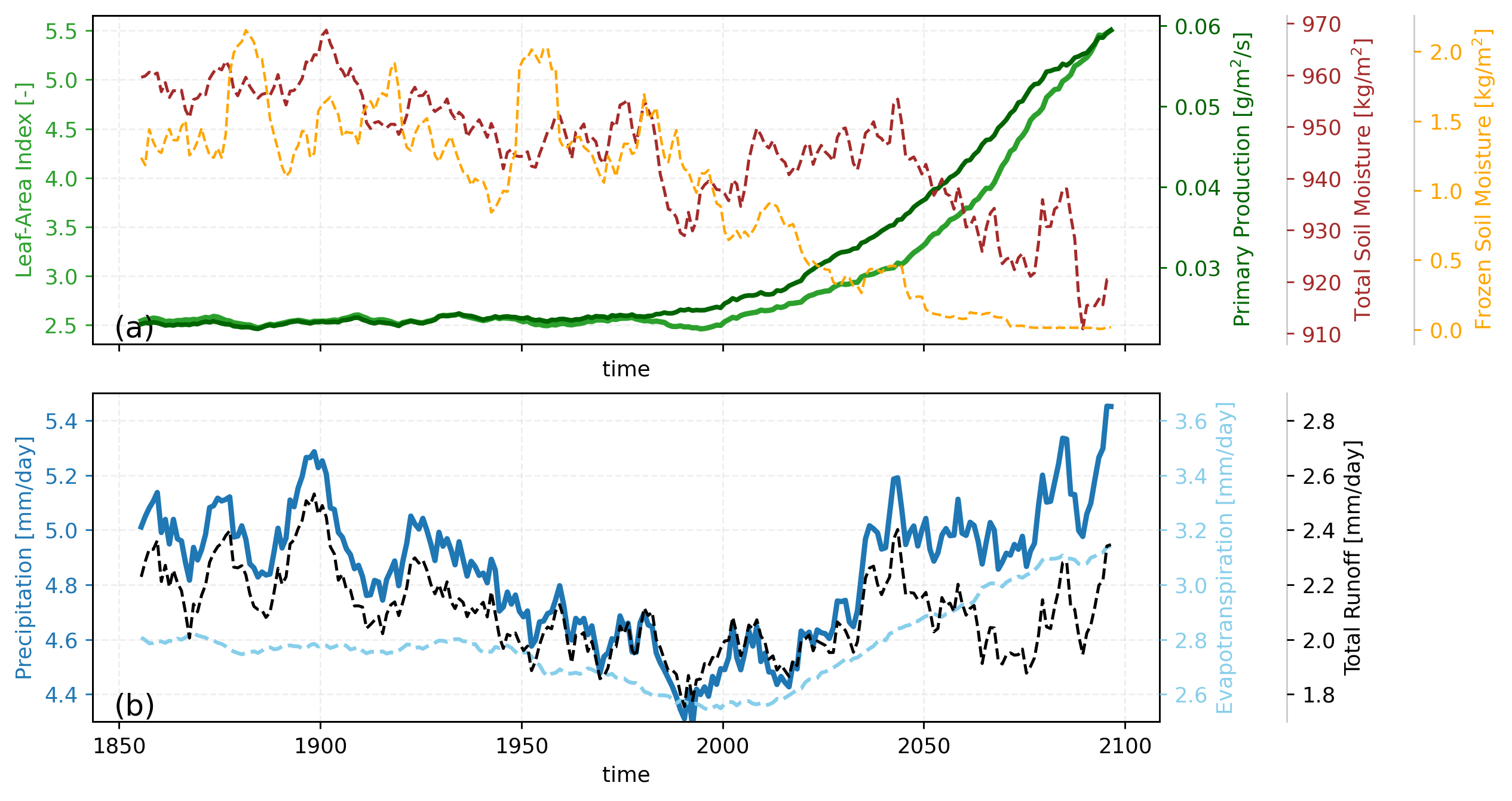}
    \caption{Time series of land and atmospheric (a,b) variables, respectively, averaged over the region associated with the gradual increase of leaf-area index in the CanESM5-CanOE model in Fig. \ref{fig: AsiaUS - increase - 1}b.}
    \label{fig: China - gradual increase - 2}
\end{figure}

\begin{figure}[H]
    \centering
    \includegraphics[width=\linewidth]{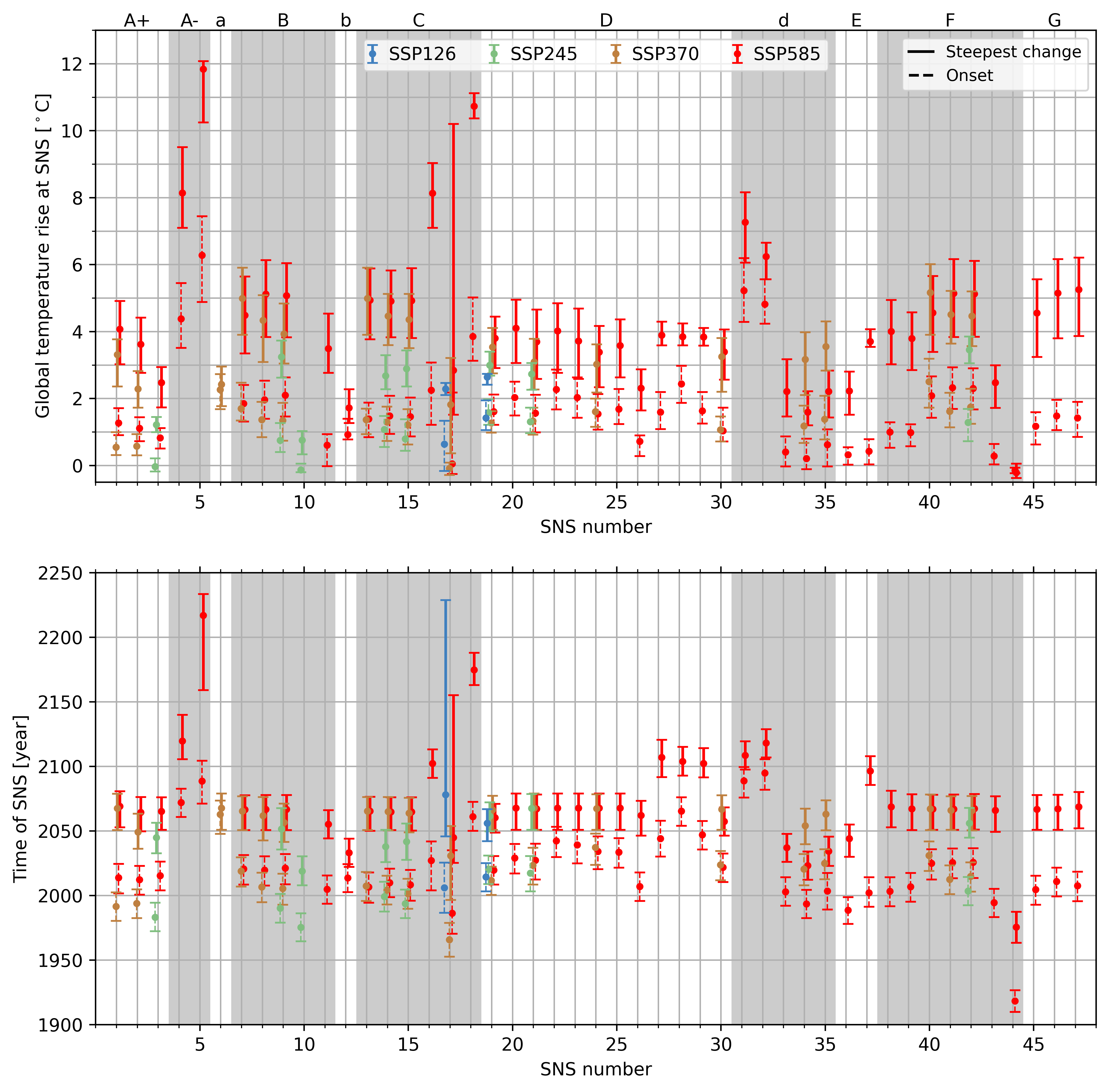}
    \caption{Global warming level distributions (upper row) and year distributions (lower row) at which SNSs begin to unfold (dashed) and most rapidly change (continuous) per each case. Dots denote the median; error bars span the central 68.2\% of the distribution (15.9th–84.1st percentile, equivalent to ±1$\sigma$ for a Gaussian distribution). The x-axis of both figures corresponds to the case number, where its models, scenarios, variables, and the fraction of members qualified as an SNS are reported in Tables~\ref{tab: all SNS cases} and \ref{tab: fraction of members giving SNS}. Table~\ref{tab: gwl per member} reports the global-warming levels of the individual variables and members associated with each case.}
    \label{fig: warming levels per case}
\end{figure}

% \appendixtables
\subsection*{Supplementary Tables}    %% Appendix A

% \begin{landscape}
\scriptsize
\renewcommand{\arraystretch}{1.1}
\setlength{\tabcolsep}{4pt}
%

\end{document}